\documentclass[twocolumn]{aastex631}

\usepackage{xspace}
\usepackage{makecell}
\usepackage{float}
\usepackage{amsmath}

\newcommand{\lya}{$\rm{Ly}\alpha$~}

\begin{document}

\title{ODIN: Confirmation and 3D Reconstruction of Six Massive Protoclusters at Cosmic Noon}

\author[0009-0008-3184-304X]{Ashley Ortiz}
\affiliation{Department of Physics and Astronomy, Purdue University, 525 Northwestern Ave., West Lafayette, IN 47906, USA}

\author[0000-0002-9176-7252]{Vandana Ramakrishnan}
\affiliation{Department of Physics and Astronomy, Purdue University, 525 Northwestern Ave., West Lafayette, IN 47906, USA}

\author[0000-0003-3004-9596]{Kyoung-Soo Lee}
\affiliation{Department of Physics and Astronomy, Purdue University, 525 Northwestern Ave., West Lafayette, IN 47906, USA}

\author[0000-0002-4928-4003]{Arjun Dey}
\affiliation{NSF NOIRLab, 950 N. Cherry Ave., Tucson, AZ 85719, USA}

\author[0000-0002-6137-0422]{Yucheng Guo}
\affiliation{School of Earth and Space Exploration, Arizona State University, Tempe, AZ 85287, USA}

\author[0009-0003-5244-3700]{Ethan Pinarski}
\affiliation{Department of Physics and Astronomy, Purdue University, 525 Northwestern Ave., West Lafayette, IN 47906, USA}

\author[0000-0001-5999-7923]{Anand Raichoor}
\affiliation{Lawrence Berkeley National Laboratory, 1 Cyclotron Road, Berkeley, CA 94720, USA}

\author[0000-0001-5567-1301]{Francisco Valdes}
\affiliation{NSF NOIRLab, 950 N. Cherry Ave., Tucson, AZ 85719, USA}


\author[0000-0003-0822-452X]{J.~Aguilar}
\affiliation{Lawrence Berkeley National Laboratory, 1 Cyclotron Road, Berkeley, CA 94720, USA}

\author[0000-0001-6098-7247]{S.~Ahlen}
\affiliation{Department of Physics, Boston University, 590 Commonwealth Avenue, Boston, MA 02215 USA}

\author[0000-0003-0570-785X]{Maria Celeste Artale}
\affiliation{Universidad Andres Bello, Facultad de Ciencias Exactas, Departamento de Fisica y Astronomia, Instituto de Astrofisica, Fernandez Concha 700, Las Condes, Santiago RM, Chile}

\author[0000-0001-9712-0006]{D.~Bianchi}
\affiliation{Dipartimento di Fisica ``Aldo Pontremoli'', Universit\`a degli Studi di Milano, Via Celoria 16, I-20133 Milano, Italy}
\affiliation{INAF-Osservatorio Astronomico di Brera, Via Brera 28, 20122 Milano, Italy}

\author[0009-0000-9758-4863]{August Bliese}
\affiliation{Department of Physics and Astronomy, Purdue University, 525 Northwestern Ave., West Lafayette, IN 47906, USA}

\author{D.~Brooks}
\affiliation{Department of Physics \& Astronomy, University College London, Gower Street, London, WC1E 6BT, UK}

\author{R.~Canning}
\affiliation{Institute of Cosmology and Gravitation, University of Portsmouth, Dennis Sciama Building, Portsmouth, PO1 3FX, UK}

\author[0009-0000-9347-1933]{Maria Candela Cerdosino}
\affiliation{Instituto de Astronomía Teórica y Experimental (IATE), CONICET-UNC, Laprida 854, X500BGR, Córdoba, Argentina}
\affiliation{Facultad de Matemática, Astronomía, Física y Computación, Universidad Nacional de Córdoba (FaMAF–UNC), Bvd. Medina Allende s/n, Ciudad Universitaria, X5000HUA, Córdoba, Argentina}

\author{T.~Claybaugh}
\affiliation{Lawrence Berkeley National Laboratory, 1 Cyclotron Road, Berkeley, CA 94720, USA}

\author[0000-0002-2169-0595]{A.~Cuceu}
\affiliation{Lawrence Berkeley National Laboratory, 1 Cyclotron Road, Berkeley, CA 94720, USA}

\author[0000-0002-1769-1640]{A.~de la Macorra}
\affiliation{Instituto de F\'{\i}sica, Universidad Nacional Aut\'{o}noma de M\'{e}xico,  Circuito de la Investigaci\'{o}n Cient\'{\i}fica, Ciudad Universitaria, Cd. de M\'{e}xico  C.~P.~04510,  M\'{e}xico}

\author{P.~Doel}
\affiliation{Department of Physics \& Astronomy, University College London, Gower Street, London, WC1E 6BT, UK}

\author[0000-0002-2890-3725]{J.~E.~Forero-Romero}
\affiliation{Departamento de F\'isica, Universidad de los Andes, Cra. 1 No. 18A-10, Edificio Ip, CP 111711, Bogot\'a, Colombia}
\affiliation{Observatorio Astron\'omico, Universidad de los Andes, Cra. 1 No. 18A-10, Edificio H, CP 111711 Bogot\'a, Colombia}

\author[0000-0003-1530-8713]{Eric Gawiser}
\affiliation{Department of Physics and Astronomy, Rutgers, the State University of New Jersey, Piscataway, NJ 08854, USA}

\author[0000-0001-9632-0815]{E.~Gaztañaga}
\affiliation{Institut d'Estudis Espacials de Catalunya (IEEC), c/ Esteve Terradas 1, Edifici RDIT, Campus PMT-UPC, 08860 Castelldefels, Spain}
\affiliation{Institute of Cosmology and Gravitation, University of Portsmouth, Dennis Sciama Building, Portsmouth, PO1 3FX, UK}
\affiliation{Institute of Space Sciences, ICE-CSIC, Campus UAB, Carrer de Can Magrans s/n, 08913 Bellaterra, Barcelona, Spain}

\author[0000-0003-3142-233X]{S.~Gontcho A Gontcho}
\affiliation{University of Virginia, Department of Astronomy, Charlottesville, VA 22904, USA}

\author[0000-0001-6842-2371]{Caryl Gronwall}
\affiliation{Institute for Gravitation and the Cosmos, The Pennsylvania State University, University Park, PA 16802, USA}
\affiliation{Department of Astronomy \& Astrophysics, The Pennsylvania State University, University Park, PA 16802, USA}

\author[0000-0002-4902-0075]{Lucia Guaita}
\affiliation{Universidad Andres Bello, Facultad de Ciencias Exactas, Departamento de Fisica y Astronomia, Instituto de Astrofisica, Fernandez Concha 700, Las Condes, Santiago RM, Chile} \affiliation{Millennium Nucleus for Galaxies (MINGAL)}

\author[0000-0003-0825-0517]{G.~Gutierrez}
\affiliation{Fermi National Accelerator Laboratory, PO Box 500, Batavia, IL 60510, USA}

\author[0000-0002-9136-9609]{H.~K.~Herrera-Alcantar}
\affiliation{Institut d'Astrophysique de Paris. 98 bis boulevard Arago. 75014 Paris, France}
\affiliation{IRFU, CEA, Universit\'{e} Paris-Saclay, F-91191 Gif-sur-Yvette, France}

\author[0000-0003-3428-7612]{Ho Seong Hwang}
\affiliation{SNU Astronomy Research Center, Seoul National University, 1 Gwanak-ro, Gwanak-gu, Seoul 08826, Republic of Korea}
\affiliation{Astronomy Program, Department of Physics and Astronomy, Seoul National University, 1 Gwanak-ro, Gwanak-gu, Seoul 08826, Republic of Korea}

\author[0000-0002-2770-808X]{Woong-Seob Jeong}
\affiliation{Korea Astronomy and Space Science Institute, 776 Daedeokdae-ro, Yuseong-gu, Daejeon 34055, Republic of Korea}

\author[0000-0003-0201-5241]{R.~Joyce}
\affiliation{NSF NOIRLab, 950 N. Cherry Ave., Tucson, AZ 85719, USA}

\author{R.~Kehoe}
\affiliation{Department of Physics, Southern Methodist University, 3215 Daniel Avenue, Dallas, TX 75275, USA}

\author[0000-0003-3510-7134]{T.~Kisner}
\affiliation{Lawrence Berkeley National Laboratory, 1 Cyclotron Road, Berkeley, CA 94720, USA}

\author[0000-0001-6356-7424]{A.~Kremin}
\affiliation{Lawrence Berkeley National Laboratory, 1 Cyclotron Road, Berkeley, CA 94720, USA}

\author[0000-0001-6270-3527]{Ankit Kumar}
\affiliation{Departamento de Ciencias Fisicas, Universidad Andres Bello, Fernandez Concha 700, Las Condes, Santiago, Chile}

\author[0000-0002-1134-9035]{O.~Lahav}
\affiliation{Department of Physics \& Astronomy, University College London, Gower Street, London, WC1E 6BT, UK}

\author[0000-0003-1838-8528]{M.~Landriau}
\affiliation{Lawrence Berkeley National Laboratory, 1 Cyclotron Road, Berkeley, CA 94720, USA}

\author[0000-0002-6810-1778]{Jaehyun Lee}
\affiliation{Korea Astronomy and Space Science Institute, 776, Daedeokdae-ro, Yuseong-gu, Daejeon 34055, Republic of Korea}

\author[0000-0001-5342-8906]{Seong-Kook Lee}
\affiliation{SNU Astronomy Research Center, Seoul National University, 1 Gwanak-ro, Gwanak-gu, Seoul 08826, Republic of Korea}

\author[0000-0001-7178-8868]{L.~Le~Guillou}
\affiliation{Sorbonne Universit\'{e}, CNRS/IN2P3, Laboratoire de Physique Nucl\'{e}aire et de Hautes Energies (LPNHE), FR-75005 Paris, France}

\author[0000-0003-4962-8934]{M.~Manera}
\affiliation{Departament de F\'{i}sica, Serra H\'{u}nter, Universitat Aut\`{o}noma de Barcelona, 08193 Bellaterra (Barcelona), Spain}
\affiliation{Institut de F\'{i}sica d’Altes Energies (IFAE), The Barcelona Institute of Science and Technology, Edifici Cn, Campus UAB, 08193, Bellaterra (Barcelona), Spain}

\author[0000-0002-1125-7384]{A.~Meisner}
\affiliation{NSF NOIRLab, 950 N. Cherry Ave., Tucson, AZ 85719, USA}

\author[0000-0002-6610-4836]{R.~Miquel}
\affiliation{Instituci\'{o} Catalana de Recerca i Estudis Avan\c{c}ats, Passeig de Llu\'{\i}s Companys, 23, 08010 Barcelona, Spain}
\affiliation{Institut de F\'{i}sica d’Altes Energies (IFAE), The Barcelona Institute of Science and Technology, Edifici Cn, Campus UAB, 08193, Bellaterra (Barcelona), Spain}

\author[0009-0008-4022-3870]{Byeongha Moon}
\affiliation{Korea Astronomy and Space Science Institute, 776 Daedeokdae-ro, Yuseong-gu, Daejeon 34055, Republic of Korea}

\author[0000-0002-2733-4559]{J.~Moustakas}
\affiliation{Department of Physics and Astronomy, Siena University, 515 Loudon Road, Loudonville, NY 12211, USA}

\author{A.~D.~Myers}
\affiliation{Department of Physics \& Astronomy, University  of Wyoming, 1000 E. University, Dept.~3905, Laramie, WY 82071, USA}

\author[0000-0001-9070-3102]{S.~Nadathur}
\affiliation{Institute of Cosmology and Gravitation, University of Portsmouth, Dennis Sciama Building, Portsmouth, PO1 3FX, UK}

\author[0000-0003-3188-784X]{N.~Palanque-Delabrouille}
\affiliation{IRFU, CEA, Universit\'{e} Paris-Saclay, F-91191 Gif-sur-Yvette, France}
\affiliation{Lawrence Berkeley National Laboratory, 1 Cyclotron Road, Berkeley, CA 94720, USA}

\author[0000-0001-9521-6397]{Changbom Park}
\affiliation{Korea Institute for Advanced Study, 85 Hoegi-ro, Dongdaemun-gu, Seoul 02455, Republic of Korea}

\author[0000-0002-0644-5727]{W.~J.~Percival}
\affiliation{Department of Physics and Astronomy, University of Waterloo, 200 University Ave W, Waterloo, ON N2L 3G1, Canada}
\affiliation{Perimeter Institute for Theoretical Physics, 31 Caroline St. North, Waterloo, ON N2L 2Y5, Canada}
\affiliation{Waterloo Centre for Astrophysics, University of Waterloo, 200 University Ave W, Waterloo, ON N2L 3G1, Canada}

\author[0000-0001-6979-0125]{I.~P\'erez-R\`afols}
\affiliation{Departament de F\'isica, EEBE, Universitat Polit\`ecnica de Catalunya, c/Eduard Maristany 10, 08930 Barcelona, Spain}

\author[0000-0001-7145-8674]{F.~Prada}
\affiliation{Instituto de Astrof\'{i}sica de Andaluc\'{i}a (CSIC), Glorieta de la Astronom\'{i}a, s/n, E-18008 Granada, Spain}

\author[0009-0006-2340-1845]{Eshwar Puvvada}
\affiliation{Department of Physics and Astronomy, Purdue University, 525 Northwestern Ave., West Lafayette, IN 47906, USA}

\author{G.~Rossi}
\affiliation{Department of Physics and Astronomy, Sejong University, 209 Neungdong-ro, Gwangjin-gu, Seoul 05006, Republic of Korea}

\author[0000-0002-9646-8198]{E.~Sanchez}
\affiliation{CIEMAT, Avenida Complutense 40, E-28040 Madrid, Spain}

\author{D.~Schlegel}
\affiliation{Lawrence Berkeley National Laboratory, 1 Cyclotron Road, Berkeley, CA 94720, USA}

\author{M.~Schubnell}
\affiliation{Department of Physics, University of Michigan, 450 Church Street, Ann Arbor, MI 48109, USA}
\affiliation{University of Michigan, 500 S. State Street, Ann Arbor, MI 48109, USA}

\author[0000-0002-3461-0320]{J.~Silber}
\affiliation{Lawrence Berkeley National Laboratory, 1 Cyclotron Road, Berkeley, CA 94720, USA}

\author[0000-0002-4362-4070]{Hyunmi Song}
\affiliation{Department of Astronomy and Space Science, Chungnam National University, 99 Daehak-ro, Yuseong-gu, Daejeon, 34134, Republic of Korea}

\author{D.~Sprayberry}
\affiliation{NSF NOIRLab, 950 N. Cherry Ave., Tucson, AZ 85719, USA}

\author[0000-0003-1704-0781]{G.~Tarl\'{e}}
\affiliation{University of Michigan, 500 S. State Street, Ann Arbor, MI 48109, USA}

\author[0000-0001-6162-3023]{Paulina Troncoso Iribarren}
\affiliation{Facultad de Ingenieria y Arquitectura, Universidad Central de Chile, Avenida Francisco de Aguirre 0405, 171-0614 La Serena, Coquimbo, Chile}

\author[0000-0001-9308-0449]{Ana Sof{\'i}a M. Uzsoy}
\affiliation{Center for Astrophysics $|$ Harvard \& Smithsonian, 60 Garden St., Cambridge, MA 02138, USA}

\author{B.~A.~Weaver}
\affiliation{NSF NOIRLab, 950 N. Cherry Ave., Tucson, AZ 85719, USA}

\author[0000-0003-3078-2763]{Yujin Yang}
\affiliation{Korea Astronomy and Space Science Institute, 776 Daedeokdae-ro, Yuseong-gu, Daejeon 34055, Republic of Korea}

\author[0000-0001-5381-4372]{R.~Zhou}
\affiliation{Lawrence Berkeley National Laboratory, 1 Cyclotron Road, Berkeley, CA 94720, USA}

\author[0000-0002-6684-3997]{H.~Zou}
\affiliation{National Astronomical Observatories, Chinese Academy of Sciences, A20 Datun Road, Chaoyang District, Beijing, 100101, P.~R.~China}



\begin{abstract}
Protoclusters represent sites of accelerated galaxy formation and extreme astrophysical activity characteristic of dense environments. Identifying massive protoclusters and mapping their spatial structures are therefore crucial for understanding how large-scale environment influences galaxy evolution. We combine wide-field Ly$\alpha$ imaging from the ODIN survey with extensive DESI and ancillary spectroscopy across the extended COSMOS and XMM-LSS fields ($\approx$14~deg$^2$) to search for massive protoclusters. We confirm six systems at $z\approx 2.4$ and $z\approx 3.1$, including three newly identified structures and three which overlap with previously known structures and/or systems detected using other tracers. We reconstruct their three-dimensional structures, estimate descendant halo masses, and, for one structure at $z\approx 3.12$, demonstrate that overlapping narrowband filters ($NB497$ and $N501$) provide accurate redshift tomography for emission-line galaxies. One protocluster at $z\approx 2.45$ overlaps with one of the LATIS tomographic fields, enabling direct comparison between galaxy and H~{\sc i} overdensities traced by Ly$\alpha$ forest absorption. Another at $z\approx 3.12$ hosts a massive quiescent galaxy ($M_{\ast} \approx 1.2 \times 10^{11}M_\odot$), suggesting that overdense environments may play a role in accelerating galaxy assembly and quenching. Comparing Ly$\alpha$ emission properties across environments, we find that protocluster galaxies exhibit higher median line fluxes and a deficit of faint emitters relative to the field. The effect is strongest when combining 2D and 3D density information, indicating that galaxies in the densest protocluster cores are most affected by environmental processes. This effect is stronger at $z\approx3.1$ than at $z\approx2.4$, suggesting possible redshift evolution. 
\end{abstract}


\section{Introduction}\label{sec:intro}

Clusters of galaxies are the most massive gravitationally bound objects in the Universe. At high redshift ($z \gtrsim 2$), the progenitors of galaxy clusters, or {\it protoclusters}, host copious star formation and AGN activity \citep[e.g.,][]{Casey2015,Umehata2015,Oteo2018}, potentially in excess of the field \citep[e.g.,][]{lemaux22,Popescu2023}. Simulations suggest that the formation of galaxies is accelerated within overdense environments \citep{Chiang2017,Muldrew2018}, supported by observations of massive and evolved galaxies in these regions \citep[e.g.,][]{Ito2023,Jin2024,tanaka24,mcconachie25,sillassen26}. Yet, the specific physical processes driving this accelerated evolution remain unclear.

The One-hundred-deg$^2$ DECam Imaging in Narrowbands \citep[ODIN:][]{Lee2024} survey has made progress towards understanding the mechanisms at play in overdense environments, by identifying a statistical sample of $\sim 150$ protoclusters at Cosmic Noon \citep{Ramakrishnan24,ramakrishnan25}. These protoclusters are identified within narrow cosmic slices ($\Delta z \sim 0.06 - 0.08$) as overdensities of Ly$\alpha$ emitting galaxies (LAEs). Analysis of their halo bias suggests that the ODIN protoclusters correspond to halos with masses of $\sim 10^{13-13.5}M_\odot$ at the time of observation, evolving into clusters with mass $\gtrsim 10^{14.5} M_\odot$ by $z = 0$ \citep{ramakrishnan25}. However, spectroscopic confirmation and detailed characterization of individual protoclusters within this large sample remain limited.

In order to establish the processes affecting galaxies in protoclusters, we must first accurately place these galaxies within the structure. The typical size of a protocluster is $\approx 5-10$~cMpc \citep{chiang13}; identifying the regions corresponding to the protocluster core, outskirts, and connected cosmic filaments thus requires mapping these regions in three dimensions with an accuracy of $\sim 2 - 3$~cMpc or better.

Recently, in \citet{ramakrishnan25b}, we presented a method to trace the 3D structures of protoclusters on scales of $\approx$50 cMpc, capturing both the most overdense peaks and the less-dense regions connecting them. This method combined LAEs photometrically selected from ODIN narrowband imaging with a subset of spectroscopically confirmed LAEs to maximize the recovery of the large-scale structure (LSS). By constructing 3D maps of two massive structures at $z \approx 3.1$, we demonstrated that extended and luminous Ly$\alpha$ nebulae or `blobs' (LABs) preferentially reside at the \emph{outskirts} of the most overdense environments. This trend was obscured by the effects of projection in 2D, highlighting the importance of a 3D perspective.

In this work, we extend the method of \citet{ramakrishnan25b} to six protocluster systems identified by ODIN, located in the COSMOS and XMM-LSS fields at $z\approx 2.4$ and $3.1$. These systems were identified as significant LAE overdensities from the ODIN surface density maps and have extensive spectroscopic data available from the Dark Energy Spectroscopic Instrument (DESI). We combine the LAEs spectroscopically confirmed by DESI with the full sample of photometrically selected ODIN LAEs to trace the 3D morphology of these structures and estimate their descendant mass. We further identify a sample of $\sim$250 LAEs which are confirmed to lie within these protoclusters and utilize them to investigate the effects of environment on the observed Ly$\alpha$ emission.

{Previous spectroscopic studies have generally been limited either by the area surveyed around protocluster candidates or by the number of spectroscopically confirmed members per system. For example, \citet{toshikawa16} identified protocluster candidates as overdensities of Lyman-break galaxies (LBGs) in $\sim$1~deg$^2$ fields and spectroscopically confirmed only a small number of systems (three protoclusters at $z\sim3\text{--}4$), each with $\sim5$ members within narrow redshift slices. As a result, these observations primarily probe the dense core regions of protoclusters. At higher redshift, \citet{Harikane2019} used narrowband-selected LAEs to identify overdense regions at $z=5.7$ and 6.6, constructing three-dimensional maps from spectroscopic samples of $\sim175$ galaxies over volumes of $\sim200\times200\times80$~cMpc$^3$, and confirming two protoclusters. While such studies have successfully confirmed individual systems—and in some cases mapped their three-dimensional structure—the limited sampling density of spectroscopic tracers prevents a full characterization of the surrounding LSS.

In contrast, our work combines wide-field photometric and spectroscopic LAE samples covering $\sim14$~deg$^2$ to reconstruct the three-dimensional density field within protocluster regions and their surrounding large-scale environments. By combining these samples, we achieve a higher tracer density within each reconstructed volume (typically $\sim 150$ spectroscopic and $\gtrsim300$ photometric LAEs in a $\sim60\times60\times60$~cMpc$^3$ region) relative to previous studies. This enables us to map the 3D density field in greater detail and to capture both the protocluster and its surrounding cosmic web.  

The structure of this paper is as follows: in Section~\ref{sec:data}, we describe the imaging and spectroscopy data, along with the selection criteria for LAEs, LABs, and protoclusters used in the analyses. In Section~\ref{sec:tomo}, we explain the approach used to estimate redshift through filter tomography and validate these results against spectroscopic redshifts within a section of one of the fields in this study. Section~\ref{sec:reconstruction} covers the 3D detection and characterization of ODIN protoclusters, and it is followed by an in-depth discussion of the six protoclusters identified using this method, in light of existing literature, in Section~\ref{sec:odin_structures}. In Section~\ref{sec:lya_properties}, we calculate the line flux of all confirmed LAEs and examine their variation across different environments. Lastly, Section~\ref{sec:summary} recaps our primary findings. We adopt a $\Lambda$CDM concordance cosmology with parameters $h = 0.7$, $\Omega_m = 0.27$, and $\Omega_\Lambda = 0.73$ and use comoving distance scales unless noted otherwise.

\begin{deluxetable*}{cccccccc}
    \tablecaption{Summary of object samples \label{tab:datasets}}
    \tablehead{\colhead{Field/Filter} & Redshift & \colhead{Effective area} & \colhead{$N_{\rm LAE}$\tablenotemark{a}} & \colhead{$N_{\rm spec}$\tablenotemark{b}} 
    & \colhead{$N_{\rm tomo}$\tablenotemark{c}}
    &\colhead{$N_{\rm LAB}$\tablenotemark{d}} & \colhead{$N_{\rm PC}$\tablenotemark{e}} \\
    & & [deg$^2$] & & & &}
    \startdata
    \hline
    COSMOS/$N419$ & 2.419--2.480 & 7.3 & 6,441 & 1,916 & - & 103 & 28 \\
    XMM-LSS/$N419$ & 2.419--2.480 & 6.6 & 5,728 & 1,161 & - & 47 & 36 \\
    COSMOS/$N501$ & 3.093--3.155 & 7.3 & 6,069 & 1,418 & - & 112 & 41 \\
    XMM-LSS/$N501$ & 3.093--3.155 & 6.6 & 3,991 & 90 & 64 & 25 & 35 \\ 
    \hline
    Total & & 13.9 & 22,229 & 4,585 & 64 & 287 & 140\\
    \hline
    \enddata
    \tablenotetext{a}{Number of photometrically selected LAEs \citep{Firestone2024}}
    \tablenotetext{b}{Number of spectroscopically confirmed LAEs \citep{pinarski2026, Guo2020}}
    \tablenotetext{c}{Number of LAEs with redshifts from dual-narrowband tomography (see Section \ref{sec:tomo})}
    \tablenotetext{d}{Number of photometrically selected LABs \citep{moon2026b}}
    \tablenotetext{e}{Number of 2D protocluster candidates \citep{Ramakrishnan24}}
\end{deluxetable*}

\section{Data} \label{data}\label{sec:data}

This study uses both photometric and spectroscopic samples of LAEs identified in the extended COSMOS and XMM-LSS fields,  covering a total sky area of $\approx$14~deg$^2$. The imaging data are taken as part of the ODIN survey \citep{Lee2024}, which covers $\approx$90~deg$^2$ in total. The selection procedures of LAEs and protoclusters, as well as spectroscopic observations, are presented in detail elsewhere \citep{Firestone2024, pinarski2026, deyInPrep}. Here, we only provide a summary of the relevant datasets. Table~\ref{tab:datasets} lists the LAE-, LAB-, and protocluster samples we utilize in this work.

\subsection{Samples of LAEs, LABs, and Protoclusters} \label{sec:odin_data}

The LAEs were initially selected based on their narrow-to-broad-band flux excess in either the $N419$ or $N501$ filters, corresponding to redshifts of $z\approx2.4$ and $z\approx3.1$, respectively. Both filters have a full width at half maximum (FWHM) of 70--75~\AA, which defines a redshift slice of $\Delta z \approx 0.06$ across each filter. A full description of the ODIN LAE photometric selection method is provided in \citet{Firestone2024}. The LAE selection criteria are applied uniformly across both the COSMOS and XMM-LSS fields, with identical flux and equivalent-width criteria for a given filter. The resulting photometric LAE counts are 6,441 ($z\approx2.4$) and 6,069 ($z\approx3.1$) in the COSMOS field, and 5,728 ($z\approx2.4$) and 3,991 ($z\approx3.1$) in the XMM-LSS field. The lower number of LAEs in the XMM-LSS field at $z\approx3.1$ is attributable to shallower imaging depth and poorer seeing, resulting in a higher 5$\sigma$ flux limit and therefore fewer detected LAEs. 

Follow-up spectroscopy indicates that contamination\footnote{The contamination fraction is defined here as the number of confirmed interlopers divided by the total number of sources that yielded secure spectroscopic redshifts.} from low-redshift interlopers is minimal—less than 4\% for $z\approx3.1$ and 6\% for $z\approx2.4$ \citep{pinarski2026}. These interlopers are roughly uniformly distributed and primarily act as a background population. As a result, they may slightly reduce the contrast of measured overdensities but do not generate artificial structures or significantly bias the identification of protoclusters. It is challenging to compare the contamination rate of the ODIN LAE samples with those reported in existing studies \citep[e.g.,][]{sobral18}, as it depends on the equivalent width cut and line flux limits, among other factors. The details of the interloper fraction as a function of NB brightness are presented in \citet{pinarski2026}.

We also utilize the LABs identified from the ODIN survey data. The selection method of LABs is presented in \citet{moon2026b}. While similar in concept to the LAE selection, it further requires a flux-dependent minimum size threshold, effectively isolating true extended sources. LABs are selected as Ly$\alpha$-emitting sources whose sizes are significantly more extended than those of point sources at a given luminosity, typically exceeding $\sim$20~arcsec$^2$. Their rest-frame equivalent width cut is similar to that for LAEs ($EW_0 >$ 20~\AA).
112 (47) and 103 (25) LABs are found in COSMOS (XMM-LSS) at $z\approx 2.4$ and 3.1, respectively. The first clustering analysis of ODIN LABs suggests that cosmic variance is non-negligible even in a relatively large survey area \citep[$> 6$~deg$^2$;][]{moon2026}.

In this work, the term 2D protocluster candidates (hereafter, 2D protoclusters) refers specifically to the projected LAE overdensity selection described below. 2D protoclusters are first identified as regions with elevated LAE surface densities, since contamination from foreground and background galaxies is expected to be minimal. Specifically, we construct surface density maps using a Voronoi tessellation of the LAE distribution, which are subsequently smoothed to suppress shot noise \citep{Ramakrishnan24}. The projected overdensity is defined as $\delta_{\rm LAE,2D} \equiv \Sigma_{\rm LAE} /\langle\Sigma_{\rm LAE}\rangle - 1$, where $\Sigma_{\rm LAE}$ is the local LAE surface density and $\langle\Sigma_{\rm LAE}\rangle$ is the field average. We then define 2D protoclusters as contiguous regions with projected overdensity $\delta_{\rm LAE,2D} > 2.3$ (2.6) at $z\approx 2.4$ (3.1) and projected area $> 40$ cMpc$^2$ \citep{ramakrishnan25}. The sky positions and angular extent of the individual 2D protocluster candidates are shown in Figure~\ref{fig:all_structures} as yellow contours. These 2D protoclusters are distinct from the 3D protoclusters that we select later in this work; as discussed in Section~\ref{sec:reconstruction}, the latter are identified in their full spatial context and
may encompass one or more 2D-selected structures. The six systems presented in this work are highlighted by the red boxes in Figure~\ref{fig:all_structures}.


\begin{figure*}
    \centering
    \includegraphics[width=\linewidth]{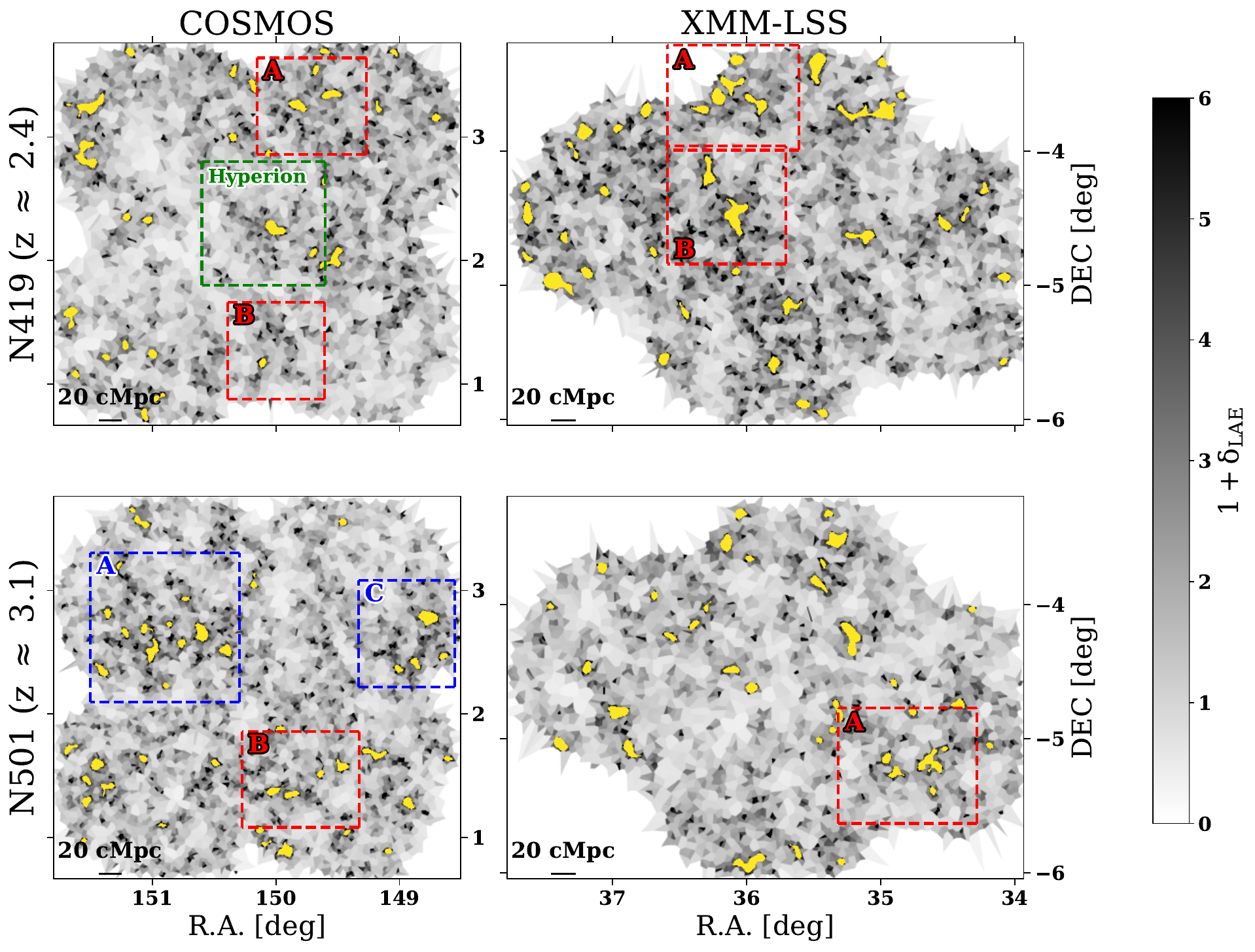}
    \caption{Protocluster candidates in COSMOS and XMM-LSS fields (yellow swathes) with the areas we consider in this work highlighted by red boxes. The greyscale background indicates the surface density of photometrically selected LAEs, with darker colors corresponding to higher values, as indicated by the colorbar. The structures considered in \citet{ramakrishnan25b} are marked in blue while a confirmed proto-super-cluster, {\it Hyperion} \citep{Cucciati2018} is marked in green.}
    \label{fig:all_structures}
\end{figure*}

\subsection{DESI-ODIN Spectroscopy} \label{sec:spec_data}

DESI is a wide-field, fiber-fed spectrograph mounted on the Mayall 4m telescope at Kitt Peak National Observatory \citep{DESI2022:instrumentation}. The instrument is currently carrying out a spectroscopic survey covering $\sim$ 17,000 deg$^2$ of the sky \citep{DESI2023:survey,DESI2025:DR1} to place constraints on the nature of dark energy \citep{DESI2025:cosmology,DESI2025:BAO}.

Based on the ODIN narrowband imaging, DESI obtained spectroscopic observations of a subset of photometric LAEs selected with less restrictive criteria than the ODIN LAEs. Unlike the ODIN LAE selection, which relies on deep Subaru broadband imaging from the Subaru Strategic Program Deep fields \citep{Aihara2019}, the DESI target selection incorporates shallower Legacy Survey broadband data to select sources over the full DESI field of view. The DESI LAE selection employs a more inclusive narrowband-to-broad-band color criterion, corresponding to a lower minimum color excess and equivalent width threshold than the ODIN LAE selection. A detailed description of this target selection will be presented in \citet{deyInPrep}.
DESI target selection is applied uniformly across the field, and the resulting spectroscopic sampling shows only minor spatial variation. These variations are not expected to introduce significant large-scale spatial inhomogeneities in the spectroscopic sampling of LAEs across the field.

DESI is capable of obtaining spectra of up to $\sim$5000 sources per pointing \citep{DESI2016:instrument_design,DESI2024:optical_corrector,DESI2024:fiber_system}, and as a result, the DESI-ODIN observations covered a total of 11,599 targets, of which 7,973 yielded reliable classifications. The redshifts of these sources were determined via visual inspection \citep{pinarski2026} of the spectra reduced via the DESI pipeline \citep{DESI2023:specz_pipeline}. The majority of sources were confirmed through the identification of the \lya line in emission, which is easily distinguishable from other emission lines due to its characteristic asymmetry, displaying a prominent red wing and suppressed blue wing.

Less than half of the DESI targets are formally selected as ODIN LAEs. However, a closer examination showed that a large fraction of the remaining sources were excluded from the ODIN LAE catalog for reasons such as falling within conservative star masks, lying outside the deep broadband coverage used by ODIN, or having narrowband excess slightly below the adopted ODIN threshold \citep[see][for a comprehensive discussion of the two LAE samples]{pinarski2026}.
Thus, while the DESI targeting includes sources selected with less restrictive criteria than the ODIN LAE selection, the resulting spectroscopically confirmed LAEs are not expected to constitute a fundamentally different population. These sources exhibit clustering properties consistent with ODIN-selected LAEs, with recent measurements showing similar galaxy bias values between the two samples \citep{white2024,herrera2025}. 

As such, we choose to utilize all DESI-confirmed LAEs classified with the spectral type\footnote{ Any source exhibiting broad lines of any kind or narrow lines typically absent in star-forming galaxies -- such as He~{\sc ii}, or C~{\sc iii}] -- is classified with the spectral type {\tt AGN}.\label{ft:spectype}} of {\tt galaxy} and quality flag\footnote{ According to the classification scheme of \citet{pinarski2026}, a quality flag of {\tt q}=4 indicates that multiple spectral features are securely detected, typically corresponding to AGN or low-redshift emission-line galaxies. A flag of {\tt q}=3 denotes the detection of a single prominent spectral feature, while {\tt q}=2 indicates a possible but uncertain feature. Given the lack of strong rest-frame UV emission lines beyond Ly$\alpha$, the vast majority of LAEs receive quality flags of {\tt q}=3 or 2. The final quality flag assigned to each source is obtained by averaging 2--5 independent evaluations.} {\tt q} $\geq 2.5$, which lie at the intended redshift range. Both quantities are determined as part of the visual inspection campaign \citep{pinarski2026,deyInPrep}, with redshifts defined by the \lya emission-line peak, and no correction to systemic redshifts applied. For the LAEs used in this work, the fitted \lya emission lines have a median signal-to-noise of $\sim7.5$.

The number of spec-z sources satisfying these criteria and subsequently used in this study is 1,916 ($z\approx 2.4$) and 1,418 ($z\approx 3.1$) in the COSMOS field and 1,161 ($z\approx 2.4$) in the XMM-LSS field, respectively. No $z\approx3.1$ LAE targets were observed in XMM-LSS with DESI (but see Section~\ref{sec:tomo} for further details). 
Although AGN host galaxies can also reside in protocluster environments at our redshift range, we exclude objects classified as \texttt{AGN}\footref{ft:spectype}, sources exhibiting broad emission lines or high-ionization features (e.g., He~{\sc ii}, C~{\sc iii}], or C~{\sc iv}), to maintain consistency with our later analysis of \lya luminosity versus environment in Section~\ref{sec:lya_properties}. The number of AGN in our spectroscopic sample at the intended redshift range is small ($<50$), so their inclusion would not affect our protocluster reconstruction \citep[see also][]{pinarski2026}.




\newcommand{\civlamb}{\ion{C}{4}\,$\lambda$1548,\,1550\xspace}
\newcommand{\heiilamb}{\ion{He}{2}\,$\lambda$1640\xspace}

We note that, while additional targeted spectroscopy has been obtained in these fields using the Keck and Gemini telescopes, those observations are concentrated primarily on the two protocluster complexes analyzed by \citet{ramakrishnan25b}, indicated by the blue boxes in Figure~\ref{fig:all_structures}. We do not include these data in our analysis because their depth and targeted sampling strategy differ substantially from the uniform DESI-ODIN spectroscopic selection. 
For a detailed discussion of these two complexes and the associated follow-up spectroscopy, we refer the reader to \citet{ramakrishnan25b}.

\section{Estimating Redshifts with Two Overlapping narrowband Filters}\label{sec:tomo}

\defcitealias{Guo2020}{G20}

\begin{figure*}
    \centering
    \includegraphics[width=0.8\linewidth]{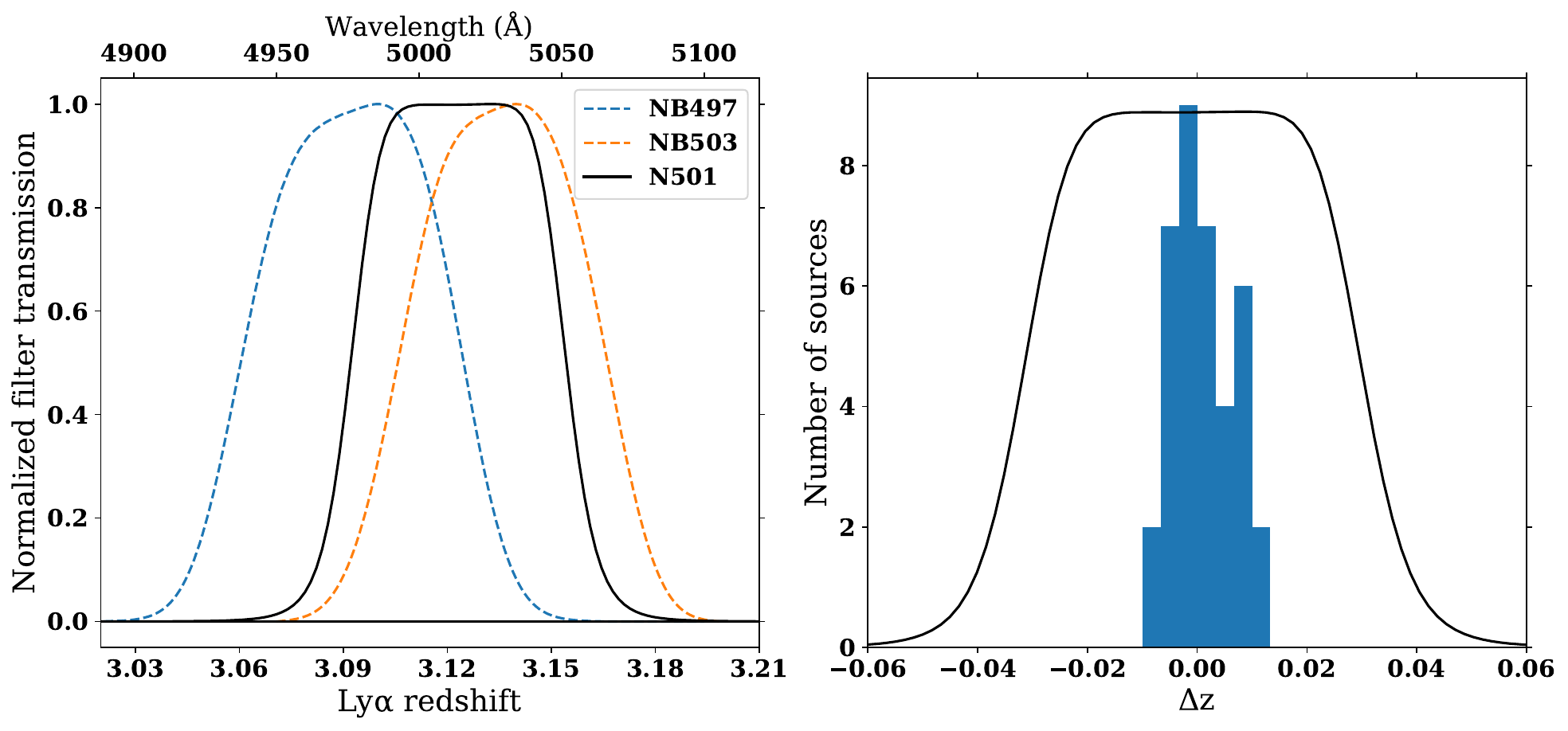}
    \caption{\emph{Left:} $NB497$ and $NB503$ filters of \citetalias{Guo2020}, shown in comparison to the ODIN $N501$ filter. The three filters have significant overlap. \emph{Right:} the offset between spectroscopic and tomographic redshifts, $\Delta z = z_{\rm spec} - z_{\rm tomo}$, is plotted together with the $N501$ transmission. The median value of offset is 0, and its standard deviation is 0.005, an order of magnitude smaller than the FWHM of the $N501$ filter.}
    \label{fig:nb_tomo}
\end{figure*}

While DESI spectroscopic coverage does not extend to $N501$ LAEs in the XMM-LSS field, a $\approx$0.7~deg$^2$ subsection of the field \citep[corresponding to the SXDS field,][]{Furusawa2008:SXDS} was imaged with two narrowband filters, $NB497$ and $NB503$ \citep[$\lambda_{\rm C}=4986$~\AA\ and 5030~\AA; FWHM $\Delta \lambda = 78$~\AA\ and 74~\AA, respectively;][]{Guo2020}. As illustrated in the left panel of Figure~\ref{fig:nb_tomo}, these filters are similar in width and overlap with $N501$. 

By combining these narrowband observations with broadband data, \citet[][hereafter G20]{Guo2020} defined LAE samples based on the $NB497$ and $NB503$ flux excess. A subset of those LAEs were followed up with MMT Hectospec observations, yielding 166 unique source redshifts. These include about 30 confirmed galaxies in and around a significant overdensity of $NB497$ LAEs at $z\approx 3.09$ (see their Figure~1). The same region is identified as a high overdensity of $N501$ LAEs by the ODIN survey.


We cross-match our $N501$ ODIN LAE catalog against the list of spectroscopically confirmed LAEs from \citetalias{Guo2020} (see their Table~2), finding 90 matches. 
Taking advantage of the availability of a large spectroscopic sample and the substantial overlap between $NB497$ and $N501$, we apply the methodology presented in \citet{Zabl:2016}, referred to as {\it narrowband tomography}, to estimate redshifts of additional sources in this overdense structure. We do not use $NB503$ for the subsequent analysis as it overlaps too much with $N501$ to be useful for tomographic purposes.


The observed flux density of a Ly$\alpha$ emitting galaxy\footnote{While the line is assumed to be Ly$\alpha$, we note that the methodology can be applied to any strong line.} at redshift $z$ observed with a narrowband filter is:
\begin{equation} \label{eq:nb_flux}
    f_{\nu,{\rm NB}} = \frac{\int \lambda f_\lambda(\lambda) \mathcal{R}(\lambda) d\lambda}{\int c \mathcal{R}(\lambda) {\lambda}^{-1} d\lambda}
\end{equation}
where $f_\lambda(\lambda)$ is the underlying spectral energy distribution of the galaxy at wavelength $\lambda$ (in units of erg~s~$^{-1}$~cm$^{-2}$~\AA$^{-1}$), $\mathcal{R}(\lambda)$ is the filter transmission, and $c$ is the speed of light (in units of cm~s$^{-1}$). We make a simplifying assumption that the flux sampled by the filter is dominated by the emission line. Additionally, we assume that the line profile can be reasonably approximated by a Dirac delta function: i.e., $f_\lambda (\lambda) \approx f_{{\rm Ly}\alpha}\delta(\lambda_0)$, where $\lambda_0=1215.67(1+z)$~\AA. In this case, Equation~\ref{eq:nb_flux} is reduced to $f_{\nu,{\rm NB}}\approx F_{{\rm Ly}\alpha}\mathcal{R}(\lambda_0)/\mathcal{C}$, where $F_{{\rm Ly}\alpha} = \lambda_0 f_{{\rm Ly}\alpha}$ is the total flux of the Ly$\alpha$ line and $\mathcal{C}$ is a filter-dependent constant.
This is justified because the intrinsic \lya line widths (FWHM $\sim 3-6$~\AA) of LAEs are extremely narrow compared to the width of our narrowband filters \citep{erb14,blaizot23}.


If the Ly$\alpha$ line falls into two adjacent narrowbands (say 1 and 2), the simplified Equation~\ref{eq:nb_flux} can be written for each filter as:
%
\begin{eqnarray}
        f_{\nu,1} &=& F_{{\rm Ly}\alpha}\mathcal{R}_{1}(\lambda_0)/\mathcal{C}_1 \nonumber 
        \\
        f_{\nu,2} &=& F_{{\rm Ly}\alpha}\mathcal{R}_2(\lambda_0)/\mathcal{C}_2
\end{eqnarray}
where ${\mathcal R}_1$ and ${\mathcal R}_2$ are the transmissions of filter 1 and 2 and $\mathcal{C}_1$ and $\mathcal{C}_2$ are the integrals over these transmission functions. As the same line is sampled by two different filters modulated by their respective transmission curves, it has a different apparent brightness in each filter. Provided that the filter transmissions are accurately measured, solving these two equations simultaneously determines the redshift of the source. 

%

We utilize this method on 36 spectroscopic sources for which $NB497$ and $N501$ fluxes have been recorded. In the right panel of Figure~\ref{fig:nb_tomo}, $\Delta z\equiv z_{\rm spec}-z_{\rm tomo}$ is depicted, where $z_{\rm tomo}$ represents the tomographic redshift estimate. The median shift is 0, with a standard deviation of $\sigma ({\Delta z})=0.005$, which is over ten times smaller than the FWHM of either filter. The dispersion equates to $\lesssim 1.2$~pMpc or 4.8~cMpc along the line-of-sight. Despite this uncertainty being greater than that of spectroscopic redshifts, it remains smaller than the average protocluster size \citep[e.g.,][]{chiang13}. Therefore, tomographic redshifts could be advantageous in analyzing the spatial structure of protoclusters and their environment. Using this methodology, we acquire redshift estimates for an additional 64 ODIN LAEs that lack spectroscopic validation. These redshift estimates are used together with spectroscopic redshifts in the 3D detection and reconstruction of ODIN protoclusters outlined in the next section. 


\section{Characterizing Spatial Structures of protoclusters} \label{sec:reconstruction}

\subsection{Probabilistic 3D Reconstruction}\label{subsec:reconstruction}

A comprehensive description of our probabilistic reconstruction technique is given in \citet[][see Section 3.1]{ramakrishnan25b}, along with thorough validation using cosmological hydrodynamic simulations processed to emulate the ODIN data \citep[refer also to,][]{Ramakrishnan24}. 
In this section, we provide a summary. 

Our method relies on the premise that spectroscopic redshifts from a randomly chosen subset of the base sample can adequately represent the overall redshift distribution within a given sightline. In the absence of LSS, the observed redshift distribution of LAEs would be expected to approximately trace the redshift selection function imposed by the narrowband filter transmission as shown in Figure~4 of \citet[][]{pinarski2026}. However, in the presence of significant overdensities, such as protoclusters, the intrinsic galaxy distribution is expected to deviate from the filter profile, reflecting real structure along the line-of-sight. 

Given the relatively narrow redshift window of our filter ($\Delta z \approx 0.06$, corresponding to the line-of-sight thickness of $\sim$60~cMpc), contamination from foreground and background interlopers is minimal, and all selected LAEs are confined to this limited comoving volume. The primary uncertainty is therefore not whether galaxies lie within the window, but rather their precise positions within it. In this context, the spectroscopic subsample provides an empirical estimate of the underlying redshift distribution, which we use as a prior for the full sample. While this assumption may break down in average or low-density sightlines -- where the galaxy distribution more closely follows the filter transmission, such that the spectroscopic samples provide limited additional information beyond the selection function -- it is better justified in high-density protocluster environments where galaxies trace underlying matter overdensities. This expectation is supported by our validation using simulations \citep[see Section~3.2 of][]{ramakrishnan25b}. 

Based on this premise, we utilize all LAEs with measured redshifts (spectroscopic and tomographic) to create sightline-dependent redshift priors. These priors are then used to statistically assign redshifts to the remaining photometric LAEs, such that the final reconstruction incorporates the full LAE sample rather than only spectroscopically confirmed sources. Given the larger uncertainties associated with tomographic redshifts relative to spectroscopic ones, we examine their effect on the reconstruction in Section~\ref{sec:redshift}.

We proceed as follows. First, we divide each protocluster volume into a three-dimensional grid with 2~cMpc spacing, equivalent to 60\arcsec--70\arcsec\ in the transverse direction and $\Delta z \approx 0.002$ along the line-of-sight. Spec-z LAEs are binned into this grid, and their distribution is smoothed using a 3D Gaussian kernel with FWHM 5--7~cMpc, to minimize shot noise while preserving spatial characteristics. A small constant is added to each bin to prevent zero-probability regions. This smoothed histogram serves as a redshift prior for each sightline. Second, by drawing from these priors, we assign redshifts for unconfirmed LAEs. In cases where no spectroscopic data is available in a given sightline, the redshift prior is assumed to be that inferred from the NB filter transmission. These unconfirmed sources are then combined with the spectroscopic LAEs, which are included in the reconstruction using their measured redshifts, such that the final density field includes \textit{all} LAEs. This process is repeated and averaged over 500 realizations, resulting in a statistically reliable 3D density field of LAEs that maps the underlying LSS at a 2~cMpc resolution.

In \citet{ramakrishnan25b}, we found that reconstructed 3D maps produced as described above give a reasonable approximation to the underlying dark matter distribution, consistently outperforming that constructed solely based on spectroscopic sources. We further showed in that work (Appendix B therein) that the peculiar motion of galaxies has minimal effect on our 3D reconstructions, which we attribute to the fact that the uncertainty in the line-of-sight position arising from such motion ($\sim 2.5 - 4$~cMpc) is much smaller than the typical size of a protocluster ($\sim 10-20$~cMpc). Similarly, the 3D reconstructions are not significantly affected by the uncertainty in the redshift estimates obtained via narrowband tomography, as we discuss in the following section. Unsurprisingly, the main limitation is set by the spectroscopic sampling fraction ($f_{\rm spec}$): i.e., the fraction of sources with spectroscopic coverage. Since these sources serve as our priors, a higher $f_{\rm spec}$ enables a more accurate recovery of the underlying dark matter distribution. Increasing $f_{\rm spec}$ from 0.2 to 1 improved the reconstruction accuracy by up to 25\%. The typical $f_{\rm spec}$ across our fields is $\sim 0.2-0.4$. 

\subsection{The Impact of Redshift Uncertainty on 3D Reconstruction}\label{sec:redshift}
The redshift information used in the 3D reconstruction is subject to several sources of uncertainty. Spectroscopic redshifts are primarily based on the peak of Ly$\alpha$ emission, which is known to be offset from the systemic velocity \citep[by 150--200~km~s$^{-1}$, with some scatter; e.g.,][]{hashimoto2013} due to radiative transfer effects. Even when systemic redshifts can be accurately measured, peculiar motion due to local gravity can shift them from true cosmological redshifts by 200-300~km~s$^{-1}$. In addition, the tomographic redshifts derived in Section~\ref{sec:tomo} generally have larger uncertainties ($\sigma_z$=0.005) than the spectroscopic redshifts.

In Appendix~B of \citet[][]{ramakrishnan25b}, we quantified how the radiative transfer effects and peculiar motion may influence the 3D reconstruction, using the IllustrisTNG300-1  simulation suite \citep{Pillepich2018a,Pillepich2018b,Nelson2019}. In a (60~cMpc)$^3$ volume centered on the $z=3$ progenitors of massive clusters, we populate a sample of mock LAEs with the number density matched to the real data. A fraction of these sources are designated as the `spectroscopic sources'; their cosmological redshifts are then perturbed accordingly to match these systematic effects. The remainder are treated as the photometric LAE sample. The 3D reconstruction procedure is then conducted in an identical manner to the real data, as described in Section~\ref{subsec:reconstruction}.

The resultant 3D model is then compared with the dark matter distribution using a metric called total variation distance \citep[TVD;][]{Tsybakov_2009}, measuring the similarity of two probability distributions. The TVD ranges from 0 for identical distributions to 1 for completely distinct distributions. We found that the difference in TVD value with or without these effects is smaller than the scatter among different realizations \citep[see Figure~15 of][]{ramakrishnan25b}.



In this work, we carry out a similar test to check the robustness of 3D reconstruction using tomographic redshifts. This time, we perturb the redshifts by drawing new values from a normal distribution centered on the true redshift and with a standard deviation of $\sigma_z=0.005$ (see Figure~\ref{fig:nb_tomo}). We find that the reconstruction using the `perturbed' redshifts has a similar TVD to that using the `true' redshifts with a typical difference of $\lesssim 5\%$, indicating that larger uncertainties from tomographic redshifts do not significantly affect our 3D reconstructions.

\begin{deluxetable*}{cccccccc}
    \tablecaption{Summary of Six Protoclusters and their Substructures \label{tab:density_peaks}}
    \tablehead{\colhead{Structure ID} & \colhead{R.A.\tablenotemark{a}} & \colhead{Decl.\tablenotemark{a}} & \colhead{$z_{\rm peak}$\tablenotemark{a}} &\colhead{$N_{\rm spec,\,peak}$} & \colhead{$V_{\rm PC}$\tablenotemark{b}} & \colhead{$\delta_g$\tablenotemark{c}} & \colhead{$\log {M^{z=0}_{\rm est}}/{M_\odot}$\tablenotemark{d}}\\
    \colhead{} & \colhead{[degree]} & \colhead{[degree]} & \colhead{} &\colhead{} & \colhead{[10$^3$ cMpc$^3$]} & \colhead{}}
    \startdata
    \hline
    \multicolumn{8}{c}{
    \textbf{COSMOS-z3.1-B} \quad 
    \textit{Center:} (149.80$^\circ$, 1.47$^\circ$) \quad 
     $N_{\rm spec}$\tablenotemark{e}$=174$, $N_{\rm LAE}$\tablenotemark{f}$=444$
    } \\
    \hline
    COSMOS-z3.1-B1 & 149.882 $\pm$ 0.161 & 1.378 $\pm$ 0.050 & 3.126 $\pm$ 0.012 & 32 & 19.8 & 5.59 $\pm$ 0.06 & 15.2 \\
    COSMOS-z3.1-B2 & 149.475 $\pm$ 0.435 & 1.567 $\pm$ 0.036 & 3.110 $\pm$ 0.004 & 10 & 4.3 & 6.25 $\pm$ 0.11 & 14.6  \\
    \hline
    \multicolumn{8}{c}{
    \textbf{COSMOS-z2.4-A} \quad 
    \textit{Center:} (149.71$^\circ$, 3.25$^\circ$) \quad 
     $N_{\rm spec}=284$, $N_{\rm LAE}=402$
    } \\
    \hline
    COSMOS-z2.4-A1 & 149.852 $\pm$ 0.117 & 3.200 $\pm$ 0.100 & 2.452 $\pm$ 0.007 & 49 & 14.0 & 3.39 $\pm$ 0.03 & 14.9 \\
    COSMOS-z2.4-A2 & 149.515 $\pm$ 0.042 & 3.306 $\pm$ 0.050 & 2.439 $\pm$ 0.003 & 8 & 4.3 & 3.42 $\pm$ 0.03 & 14.4 \\
    \hline
    \multicolumn{8}{c}{
    \textbf{COSMOS-z2.4-B} \quad 
    \textit{Center:} (150.00$^\circ$, 1.27$^\circ$) \quad 
    $N_{\rm spec}=183$, $N_{\rm LAE}=320$
    } \\
    \hline
    COSMOS-z2.4-B1 & 149.970 $\pm$ 0.057 & 1.439 $\pm$ 0.052 & 2.456 $\pm$ 0.005 & 12 & 7.7 & 3.10 $\pm$ 0.02 & 14.6 \\
    COSMOS-z2.4-B2 & 150.035 $\pm$ 0.093 & 1.179 $\pm$ 0.055 & 2.437 $\pm$ 0.005 & 21 & 7.1 & 3.45 $\pm$ 0.04 & 14.6 \\
    \hline
    \multicolumn{8}{c}{
    \textbf{XMM-z2.4-A} \quad 
    \textit{Center:} (36.10$^\circ$, -3.60$^\circ$) \quad 
    $N_{\rm spec}=150$, $N_{\rm LAE}=377$
    } \\
    \hline
    XMM-z2.4-A1 & 36.029 $\pm$ 0.146 & -3.625 $\pm$ 0.070 & 2.428 $\pm$ 0.005 & 27 & 16.9 & 7.88 $\pm$ 0.08 & 15.2 \\
    XMM-z2.4-A2 & 36.340 $\pm$ 0.027 & -3.728 $\pm$ 0.022 & 2.441 $\pm$ 0.002 & 3 & 0.8 & 6.78 $\pm$ 0.08 & 13.8 \\
    \hline
    \multicolumn{8}{c}{
    \textbf{XMM-z2.4-B} \quad 
    \textit{Center:} (36.15$^\circ$, -4.40$^\circ$) \quad 
     $N_{\rm spec}=158$, $N_{\rm LAE}=708$
    } \\
    \hline
    XMM-z2.4-B & 36.162 $\pm$ 0.105 & -4.533 $\pm$ 0.098 & 2.450 $\pm$ 0.011 & 17 & 16.5 & 3.40 $\pm$ 0.02 & 15.0 \\
    \hline
    \multicolumn{8}{c}{
    \textbf{XMM-z3.1-A} \quad 
    \textit{Center:} (34.80$^\circ$, -5.20$^\circ$) \quad 
   $N_{\rm spec}=73$, $N_{\rm tomo}=33$, $N_{\rm LAE}=433$
    } \\
    \hline
    XMM-z3.1-A1 & 35.006 $\pm$ 0.101 & -5.159 $\pm$ 0.093 & 3.121 $\pm$ 0.006 & 7 & 17.9 & 7.13 $\pm$ 0.07 & 15.2 \\
    XMM-z3.1-A2 & 34.574 $\pm$ 0.027 & -5.216 $\pm$ 0.102 & 3.115 $\pm$ 0.006 & 19 & 15.0 & 6.43 $\pm$ 0.06 & 15.1 \\
    \hline
    \enddata
    \tablenotetext{a}{The centroid of each peak estimated from the 3D reconstruction. The uncertainties are computed as the density-weighted second moment of the positions of the voxels.}
    \tablenotetext{b}{The volume of the density peak above a density threshold of 97\%, see text}
    \tablenotemark{c}{The median 3D galaxy overdensity within the protocluster volume. The uncertainties are estimated via bootstrap resampling.}
    \tablenotetext{d}{The uncertainty in the estimated descendant mass is $\sim$ 0.2--0.3~dex.}
    \tablenotetext{e}{Total number of spectroscopically confirmed galaxies (DESI) within the 3D reconstruction.}
    \tablenotetext{f}{Total number of photometrically selected LAEs within the 3D reconstruction.}
\end{deluxetable*}

\subsection{Detection and Descendant Mass Estimates of Protoclusters}\label{sec:desc_mass}


From the reconstructed maps created as described in Section~\ref{subsec:reconstruction}, we identify density peaks as contiguous regions in which each voxel of the three-dimensional density field reaches or exceeds the 97th percentile and with a volume greater than 500 cMpc$^3$. In this work, the term 3D protoclusters refers specifically to the density peaks selected using this 3D overdensity criterion. This definition is distinct from the 2D protocluster selection described in Section~\ref{sec:odin_data}, which is based on projected LAE overdensities. The resulting enclosed volume of each 3D protocluster is defined as the protocluster volume, $V_{\rm PC}$. The minimum volume threshold of 500 cMpc$^3$ is roughly equivalent to a sphere of radius $5$ cMpc, which is the expected size of the progenitor of a $M_{200} \sim10^{14}M_{\odot}$ cluster at $z\sim2-3$ based on simulations \citep{chiang13,muldrew15}. It is also consistent with the effective line-of-sight resolution set by the $z_{\rm spec}-z_{\rm tomo}$ scatter (4.8 cMpc). The 97th percentile density threshold was calibrated using IllustrisTNG mock catalogs \citep[][Section 6.2]{ramakrishnan25b}, such that the enclosed volumes above this threshold trace the matter that will collapse into the final $z=0$ cluster, yielding descendant-mass estimates with a scatter of only 0.2--0.3~dex relative to their true values.

The 3D protoclusters identified as described above may encompass multiple 2D protoclusters identified as described in Section \ref{sec:odin_data}. This can arise for two main reasons. First, the 2D protocluster selection is based on projected LAE surface density maps, in which galaxies at different locations along the line-of-sight may contribute to the same projected overdensity. Thus, some 2D overdensities may not correspond to coherent structures in redshift space. Conversely, multiple 2D protoclusters that are separated on the sky may be connected in three dimensions if their LAEs occupy a similar redshift range and are linked by a lower density structure in the reconstructed field. Second, while the 3D reconstruction incorporates line-of-sight information to recover the underlying spatial distribution, its effective resolution is limited by the sampling density of spectroscopic redshifts. As a result, closely spaced density peaks may not be fully resolved and can instead appear as a single contiguous density peak in the reconstruction.



Descendant masses, $M_{\rm est}^{z=0},$ are estimated as the total mass within $V_{\rm PC}$ as \citep{Steidel1998,Steidel2000}:
\begin{equation}\label{eq:descendant_mass}
    M^{z=0}_{\rm est} = \left(1+\frac{\delta_{g}}{b_{g}}\right) \rho_{0}V_{\rm PC}
\end{equation}
where $\rho_{0}$ is the mean comoving cosmic density, $\delta_g$ is the median 3D galaxy overdensity within the protocluster volume, and $b_g$ is the galaxy bias. The latter two are related to matter overdensity, $\delta_m$, as: $\delta_g = b_g \delta_m$. 
Clustering measurements of ODIN LAEs give $b_g = 1.7 \pm 0.2$ and $b_g = 2.0 \pm 0.2$ at $z\approx2.4$ and $z\approx3.1$ \citep{white2024,herrera2025}. Since our analysis includes both redshifts, we could, in principle, adopt distinct bias values at each redshift; however, the difference between the two is small. Varying $b_g$ within this range changes the inferred descendant masses only minimally: using $b_g = 1.8$ instead of the upper-end value $b_g = 2.2$ modifies $\log_{10}(M_{\rm est}^{z=0})$ by $\lesssim 0.07$~dex, far below our overall uncertainty of 0.2--0.3~dex. For this reason, we adopt a single representative value $b_g = 1.8$.

Within the surveyed area of 14~deg$^2$, the DESI spectroscopy provided a relatively sparse sampling of our photometric LAEs. Furthermore, the fiber-assignment algorithm leads to non-uniform coverage, meaning some regions are more densely sampled than others. These factors affect the detectability of all protoclusters, with those having smaller angular sizes or situated near the center of DESI pointings being most impacted. Consequently, our main objective in this study is not to confirm all protoclusters in these fields but to reliably identify protoclusters that are well characterized by the DESI data. Therefore, our analysis in this work is limited to structures that show significant overdensities in the 3D reconstruction. As a result, our search is biased towards the largest and most massive protoclusters in these regions.

In this work, we confirm six protoclusters at $z\approx2.4$ and $z\approx3.1$. Three of these are newly identified systems confirmed using DESI–ODIN spectroscopy, while two correspond to previously known structures that we independently recover and confirm with the same dataset. In addition, we confirm a sixth structure previously reported by \citetalias{Guo2020} using existing spectroscopic redshifts and tomographic redshift estimates described in Section~\ref{sec:tomo}. In reconstructing its structure, we utilize both spectroscopic and tomographic redshifts to determine redshift priors for the probabilistic reconstruction. 

The key properties of the six identified structures are summarized in Table~\ref{tab:density_peaks}, including their positional centers, the numbers of photometric and spectroscopic LAEs used to create the 3D reconstruction, and the geometric center, estimated volume, and inferred descendant mass of each density peak within the structure. Their sky locations are also shown in Figure~\ref{fig:all_structures}. In all but one case, we identify two or more substructures that are spatially adjacent or interconnected; for these, we list the same parameters separately for each density peak. 
In some cases, depending on the proximity of the subgroups and the total mass contained within the system, the subgroups may merge into a single massive cluster by $z = 0$. Consequently, the summed masses of the subgroups should be regarded as a lower limit on the total descendant mass of the eventual merged system. 
We also report the number of DESI-confirmed galaxies that lie within the volume of each density peak.

Smaller substructures likely exist, as was the case for COSMOS-z3.1-A and COSMOS-z3.1-C, which were studied in greater detail with targeted spectroscopy \citep{ramakrishnan25b}. However, the generally lower spectroscopic sampling density here makes it challenging to identify smaller and/or subtler features.

Four of the structures listed in Table~\ref{tab:density_peaks} have a total mass $\log(M/M_\odot) \gtrsim 15$. Including COSMOS-z3.1-A, COSMOS-z3.1-C, and the {\it Hyperion} proto-supercluster \citep[][indicated in green in Figure \ref{fig:all_structures}]{Cucciati2018}, a total of seven such structures have been identified within the ODIN-DESI area. The total cosmic volume spanned by these data is $\sim$ $1.6 \times 10^7$~cMpc$^3$ \citep{ramakrishnan25}. Assuming a \citet{Sheth1999} halo mass function, the number density of halos with $\log(M/M_\odot) \gtrsim 15$ at $z = 0$ is expected to be $\sim 2.2 \times 10^{-7}$~cMpc$^{-3}$; we would thus expect $\approx$4 such structures within the current ODIN volume, consistent with our present results.

In \citet{ramakrishnan25b}, we defined a naming convention for spectroscopically confirmed ODIN protoclusters to be the field name, redshift, followed by a unique letter connected by a hyphen, and reported the discovery of COSMOS-z3.1-A and COSMOS-z3.1-C. We maintain this naming convention in the present work. Combined with this work, eight ODIN protoclusters have been spectroscopically confirmed. 

\begin{figure*}
    \centering
    \textbf{COSMOS-z3.1-B}\par\medskip
    \begin{minipage}[c]{0.57\linewidth}
        \includegraphics[width=\linewidth]{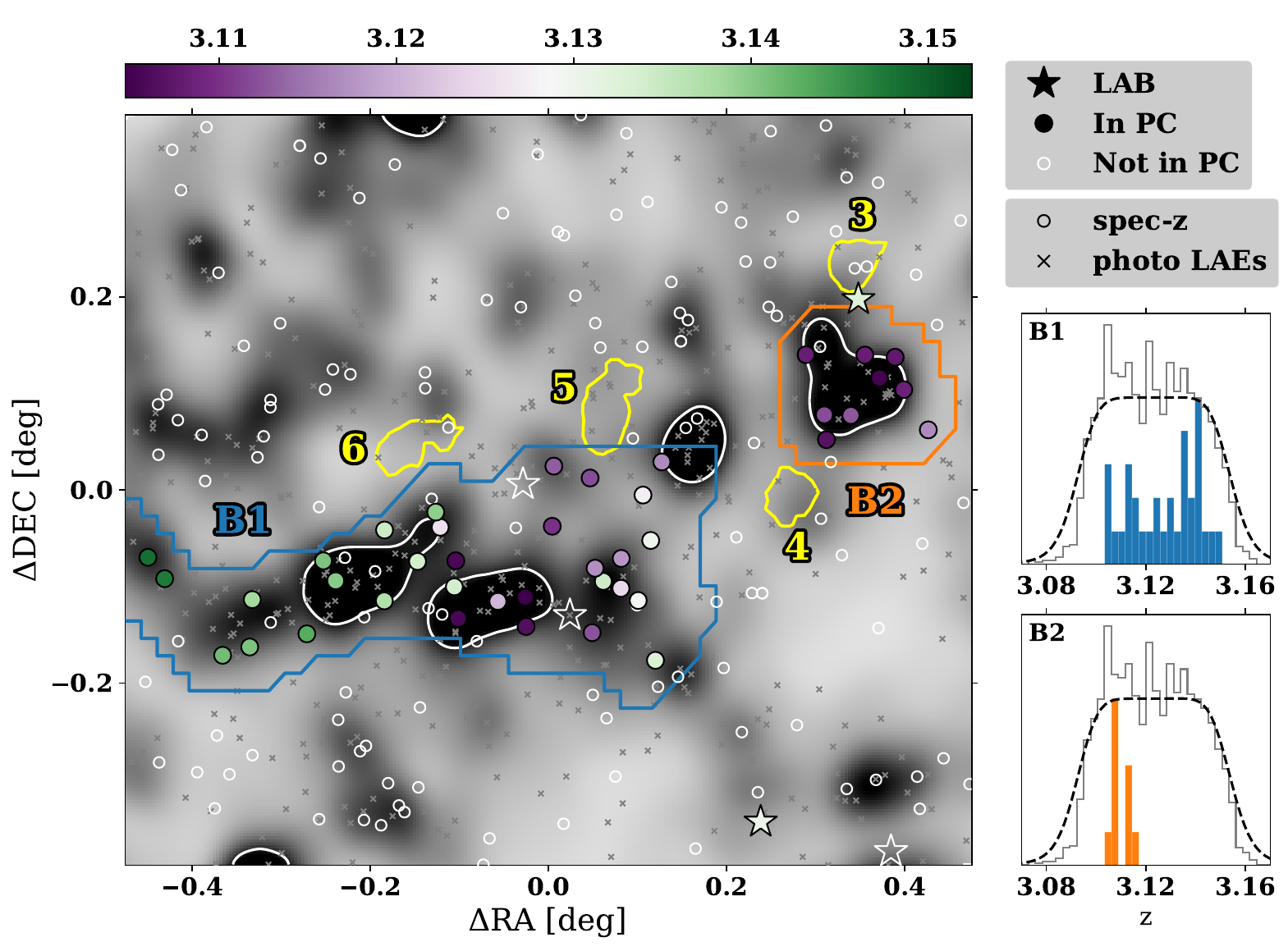}
    \end{minipage}%
    \hfill
    \begin{minipage}[c]{0.41\linewidth}
        \begin{interactive}{js}{COSMOS-z3.1-B-LAB.x3d}
            \includegraphics[width=\linewidth]{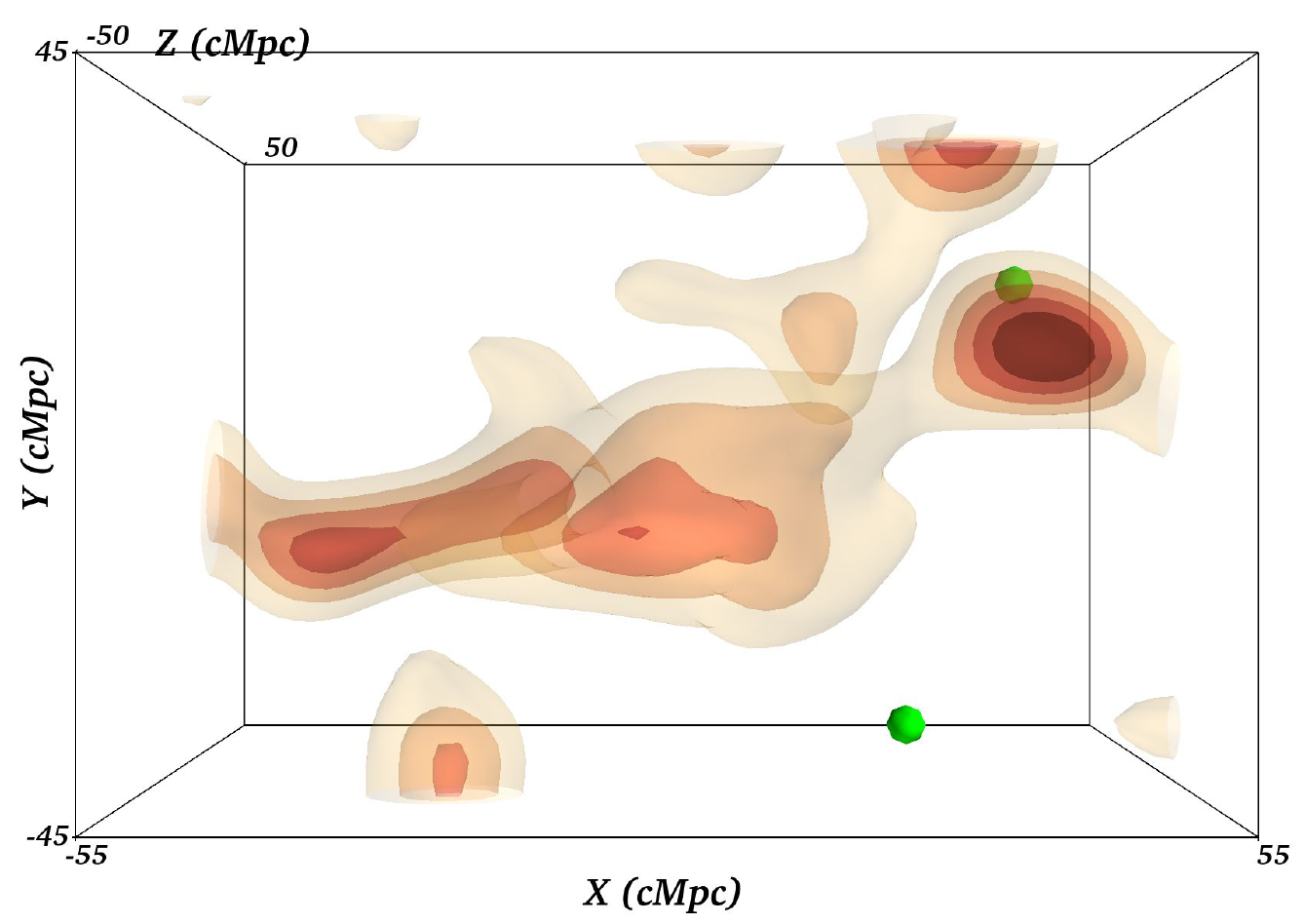}
        \end{interactive}
    \end{minipage}
    \includegraphics[width=0.96\linewidth]{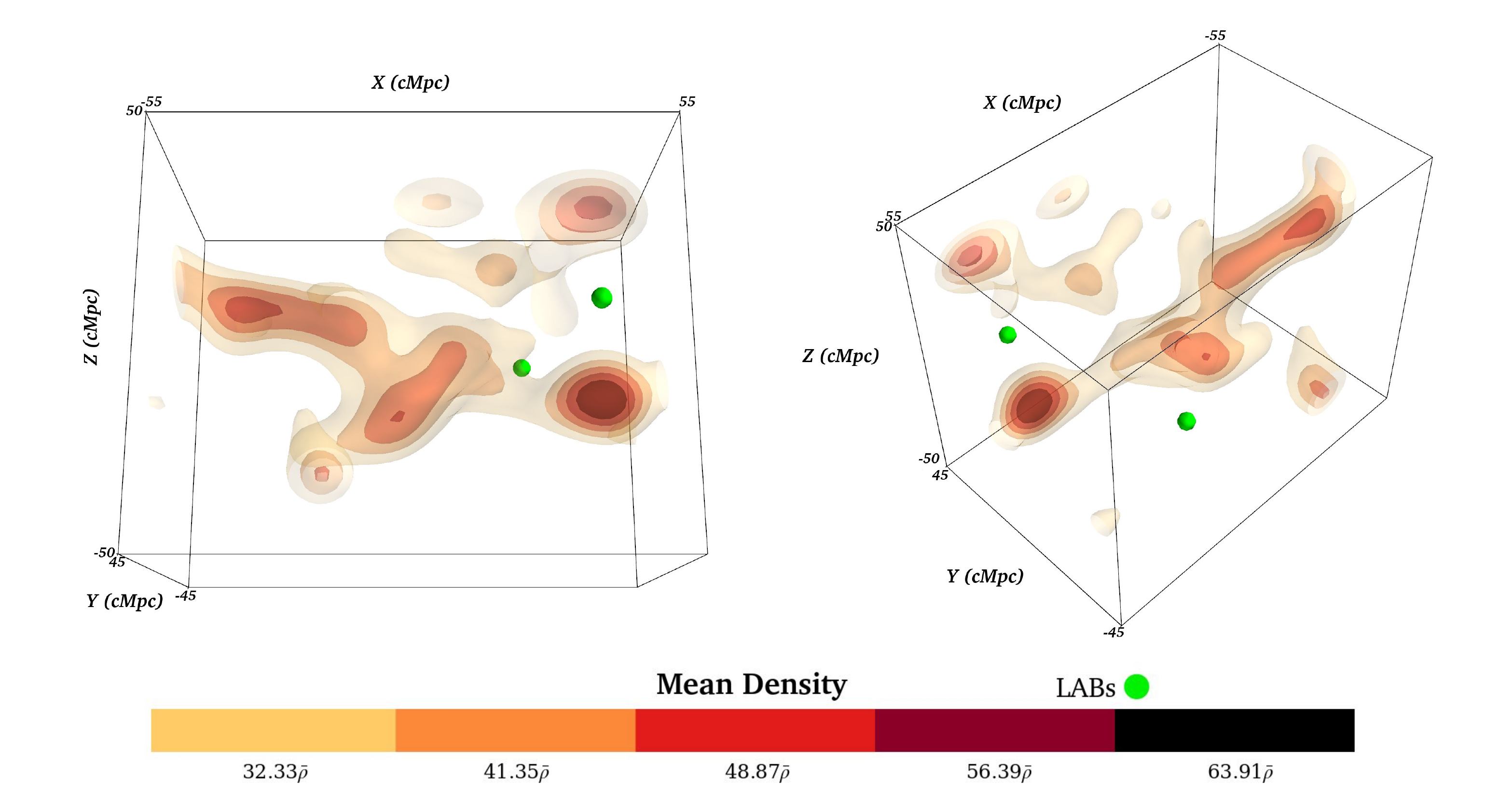}
    \caption{
    Top left: LAE surface density map around COSMOS-z3.1-B, shown in greyscale as in Figure~\ref{fig:all_structures}; the pointing center is listed in Table~\ref{tab:density_peaks}. Blue and orange contours mark the projected 3D density peaks B1 and B2, while white contours indicate 2D protocluster candidates identified from the 2D LAE overdensity map \citep{Ramakrishnan24,ramakrishnan25}. LAEs with spectroscopic redshifts in B1 and B2 are shown as filled circles, color-coded by redshift; spectroscopic sources that fall outside these 3D density peaks are shown as open circles. LABs are shown as stars, and all photometric LAEs as grey crosses. Yellow contours denote protocluster candidates identified by \citep{toshikawa24}. Side panels show the redshift distributions of confirmed galaxies in each 3D peak (colored histograms), with the overall distribution shown in grey and the filter transmission curve as a dashed line.
Top right: Reconstructed 3D model of COSMOS-z3.1-B viewed along the same line-of-sight, with densities relative to the mean field indicated by the color bar; confirmed LABs are shown as green spheres.
Bottom panels show the 3D model from two additional viewing angles. An interactive version of the 3D reconstruction is available on \href{https://ortiz140.github.io/odin/}{GitHub} and in the online article. The figures can be explored using mouse controls for panning, zooming, and rotation. Additional information about these controls is available \href{https://www.x3dom.org/documentation/interaction/}{here}.
    }
\label{fig:cosmos_z3.1_B}
\end{figure*}

\section{ODIN structures in 3D} \label{sec:odin_structures}

A key advantage of our 3D reconstruction is that it separates structures along the line-of-sight, reveals physically connected components, and traces filamentary features that are often difficult to discern in projected (2D) density maps. In this section, we examine each of the six protoclusters in greater detail, emphasizing cases where the 3D\footnote{Interactive versions of these 3D visualizations are available on GitHub (\href{https://ortiz140.github.io/odin/}{https://ortiz140.github.io/odin/}).} analysis yields insights into their structure and physical associations that cannot be inferred from projected overdensities alone. Whenever possible, we also place these systems in the context of complementary data available along their sightlines. Notably, one structure is identified as an LBG-rich protocluster candidate by another survey, another falls in a field with IGM tomography data, and the third harbors a massive ($M_{\rm star}\approx 10^{11.2}M_\odot$) quiescent galaxy.

\subsection{COSMOS-z3.1-B}
This structure was first identified by \citet{ramakrishnan23} as one of the three most significant peaks, designated as \textit{Complex B} in their study. COSMOS-z3.1-B is probably the second largest structure in this analysis (Table~\ref{tab:density_peaks}). In the 2D density map (shown as greyscale and white contours in the top left panel of Figure~\ref{fig:cosmos_z3.1_B}), the primary overdensity stretches from east to west and is distinct from two other o verdensities to the northwest. The 3D reconstruction indicates that the main overdensity is composed of two broadly connected structures (labeled `B1'), where the eastern section is more distant from us compared to the western section and the northwestern peak (`B2'). This is best illustrated in the bottom left panel, which shows the line-of-sight (Z) direction. 

Three LABs are found near COSMOS-z3.1-B, away from the highest density peaks, one of which has been confirmed by \citet{moon2026b} at $z\approx3.133$ near density peak B2. Although appearing to lie within an overdense region in projection, the blob is $\sim20$~cMpc behind the peak illustrating how the 3D reconstruction helps distinguish physically unrelated systems that would otherwise appear associated in projection.


Recently, \citet{toshikawa24} identified protocluster candidates as overdensities of Lyman break galaxies (LBGs) in the deep and ultradeep fields of the Subaru Strategic Program \citep{Aihara2022}. Given that the redshift selection of LBGs is much broader ($\Delta z\sim 0.6$) than that of LAEs ($\Delta z\sim 0.06$, corresponding to $\Delta d_{\rm LOS}\sim 60$~cMpc), only the highest density ``cores'' regions are expected to be detected. In the top left panel of Figure~\ref{fig:cosmos_z3.1_B}, we reproduce the $2\sigma$ contours of four protoclusters identified by \citet{toshikawa24} as yellow contours, labeled with their ID numbers following their Table 2. These four are among the nine protoclusters they identified within the full $\approx$8~deg$^2$ extended COSMOS field at $z \sim 2.6-3.6$.

Intriguingly, these four protocluster candidates are found within this relatively small ($\approx 1^\circ \times 0.8^\circ$) section of the field, mere arcminutes away from COSMOS-z3.1-B. Moreover, the LBG- and LAE density peaks appear to systematically avoid one another, although, if any of the former lie at the same redshift as the latter, it is expected that the true extent would be likely much larger and therefore would overlap significantly with COSMOS-z.3.1-B.

We consider possible scenarios. First, the observed angular proximity could be attributed to the projection effect, indicating that the four LBG protoclusters are not in close physical proximity to COSMOS-z.3.1-B. A high concentration of LBG-rich protoclusters may be due to cosmic variance since protoclusters show strong clustering \citep{toshikawa18,ramakrishnan25}. One of these protoclusters, \textit{ID5}, has been confirmed to be a protocluster at $z\approx3.05$ (J. Toshikawa, private communication), placing it in the foreground relative to density peak B1. Given that the line-of-sight separation between  COSMOS-z3.1-B1 and \textit{ID5} is more than 70~cMpc, they are unlikely to be associated with each other. 

Alternatively, one or more of the LBG-rich protoclusters could be genuinely associated with COSMOS-z.3.1-B. How LBGs and LAEs trace the LSS is still poorly understood. \citet{Shi2019} reported a segregation of an LAE and LBG overdensity and speculated that the latter may be dominated by more evolved and more massive galaxy constituents, perhaps due to an early formation time. 
Using cosmological simulations, \citet{im2024} demonstrated that LAEs and LBGs can occupy the same LSS while appearing spatially distinct due to selection effects and galaxy bias.
In contrast, in a few best-known protocluster systems such as SSA22 and {\it Hyperion}, LAEs and LBGs appear to coexist \citep[e.g.,][]{Cucciati2018,huang22}. Targeted spectroscopy of galaxies belonging to these regions of overdensities will be needed to obtain a clear picture.

\subsection{COSMOS-z2.4-A} 

\begin{figure*}
    \centering
    \textbf{COSMOS-z2.4-A}\par\medskip
    \begin{minipage}[c]{0.56\linewidth}
        \includegraphics[width=\linewidth]{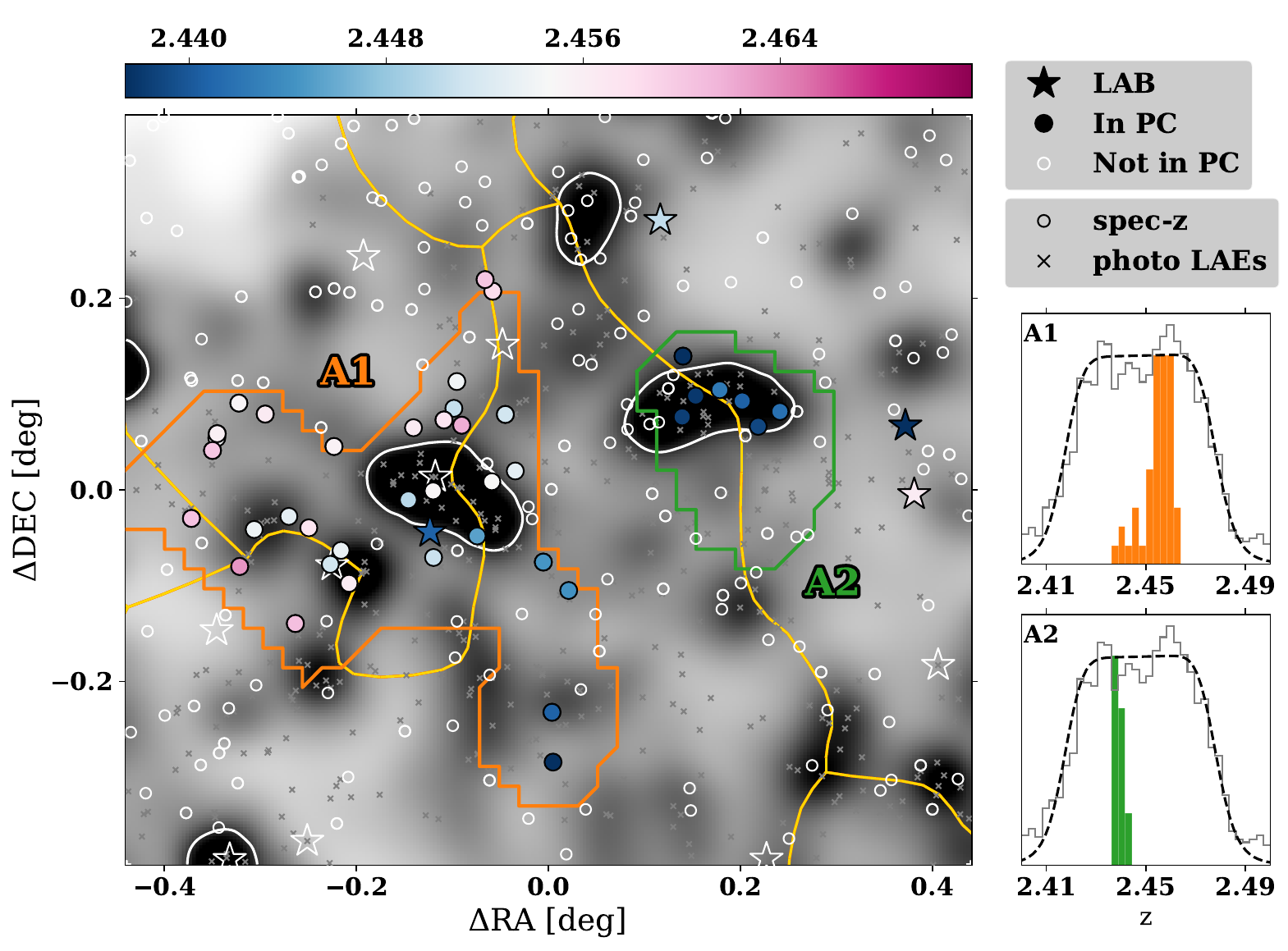}
    \end{minipage}
    \hfill
    \begin{minipage}[c]{0.40\linewidth}
        \begin{interactive}{js}{COSMOS-z2.4-A-LAB.x3d}
            \includegraphics[width=\linewidth]{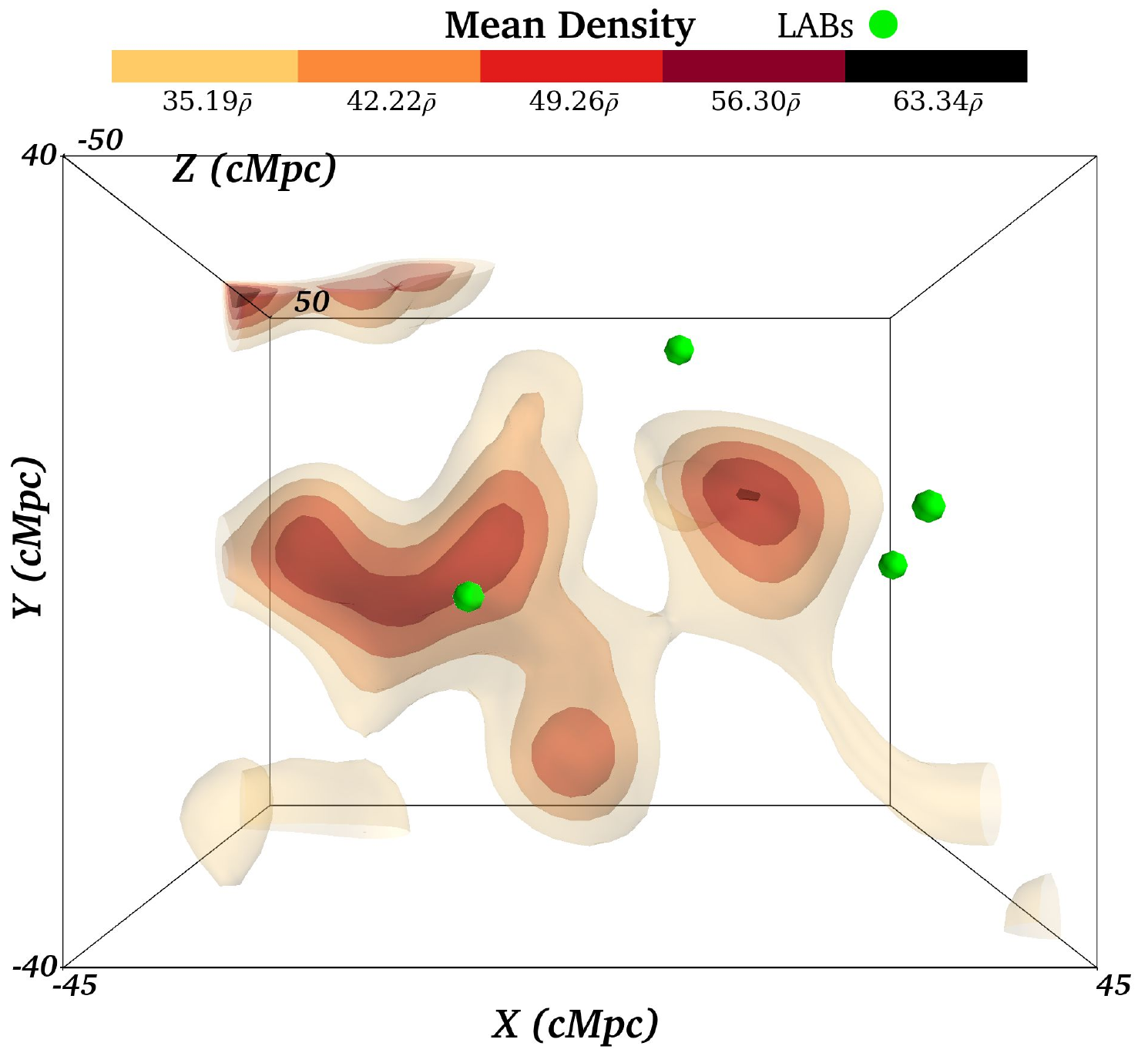}
        \end{interactive}
    \end{minipage}
    \caption{Similar to Figure~\ref{fig:cosmos_z3.1_B}, but for COSMOS-z2.4-A. The yellow line segments in the left panel show filaments identified using \texttt{DisPerSE} applied to the 2D distribution of photometric LAEs.}
    \label{fig:cosmos_z2.4_A}
\end{figure*}

Like COSMOS-z3.1-A and B, this structure is extended in the transverse (sky) direction, stretching $\approx$90$\times$80~cMpc$^2$ at $z\approx2.45$. However, it is a substantially less massive and smaller protocluster than COSMOS-z3.1-A\footnote{As described in \citet{ramakrishnan25b}, COSMOS-z3.1-A spans $\sim140 \times 140$ cMpc$^2$ in the sky direction with the descendant mass a few times larger than the present-day Coma cluster.} and COSMOS-z3.1-B. As illustrated in Figure~\ref{fig:cosmos_z2.4_A}, we have identified two main density peaks that are closely grouped in redshift space. The southern portion of the larger, more massive group `A1' ($\log M_{\rm est}^{z=0}/M_\odot \approx 14.9\pm 0.2$) curves slightly in our direction, while the smaller, less massive group `A2' lies closer to us and sits 15~cMpc from it in 3D space. 

COSMOS-z2.4-A hosts five LABs along its sightline and several more at a distance, of which four have been confirmed by DESI. Similar to the trend observed in \citet{ramakrishnan25b}, the confirmed blobs, while proximate to the 3D density peaks, seem to avoid their cores and instead reside at the outskirts. 

The 3D reconstruction reveals a filamentary extension emerging from peak A2 and stretching toward the bottom-right corner of the volume. This structure is not evident in the 2D surface density map, as filaments correspond to extended, moderate overdensities that become diluted by line-of-sight projection \citep{Ramakrishnan24}, underscoring the ability of the 3D approach to recover features that would otherwise remain hidden. Nevertheless, several spectroscopically confirmed LAEs outside the main protocluster density peaks (open circles) trace the projected path of this extension.

A consistent filamentary structure is also recovered with the filament-finding algorithm \texttt{DisPerSE} \citep{Sousbie2011a}, which identifies a network of connected segments in both the 2D distribution of photometric LAEs (shown in yellow) and in 3D using the sky positions and redshifts of confirmed LAEs. The reliability of filament detections near massive protoclusters has been demonstrated with IllustrisTNG simulations in \citet{Ramakrishnan24}. Although the limited spectroscopic sampling precludes a fully quantitative assessment of its statistical significance, its consistent appearance in the 3D reconstruction, spectroscopic tracers, and \texttt{DisPerSE} results strongly suggests that it is a genuine feature physically associated with COSMOS-z2.4-A2.


\subsection{COSMOS-z2.4-B}

\begin{figure*}
    \centering
    \textbf{COSMOS-z2.4-B}\par\medskip
    \begin{minipage}[c]{0.56\linewidth}
        \includegraphics[width=\linewidth]{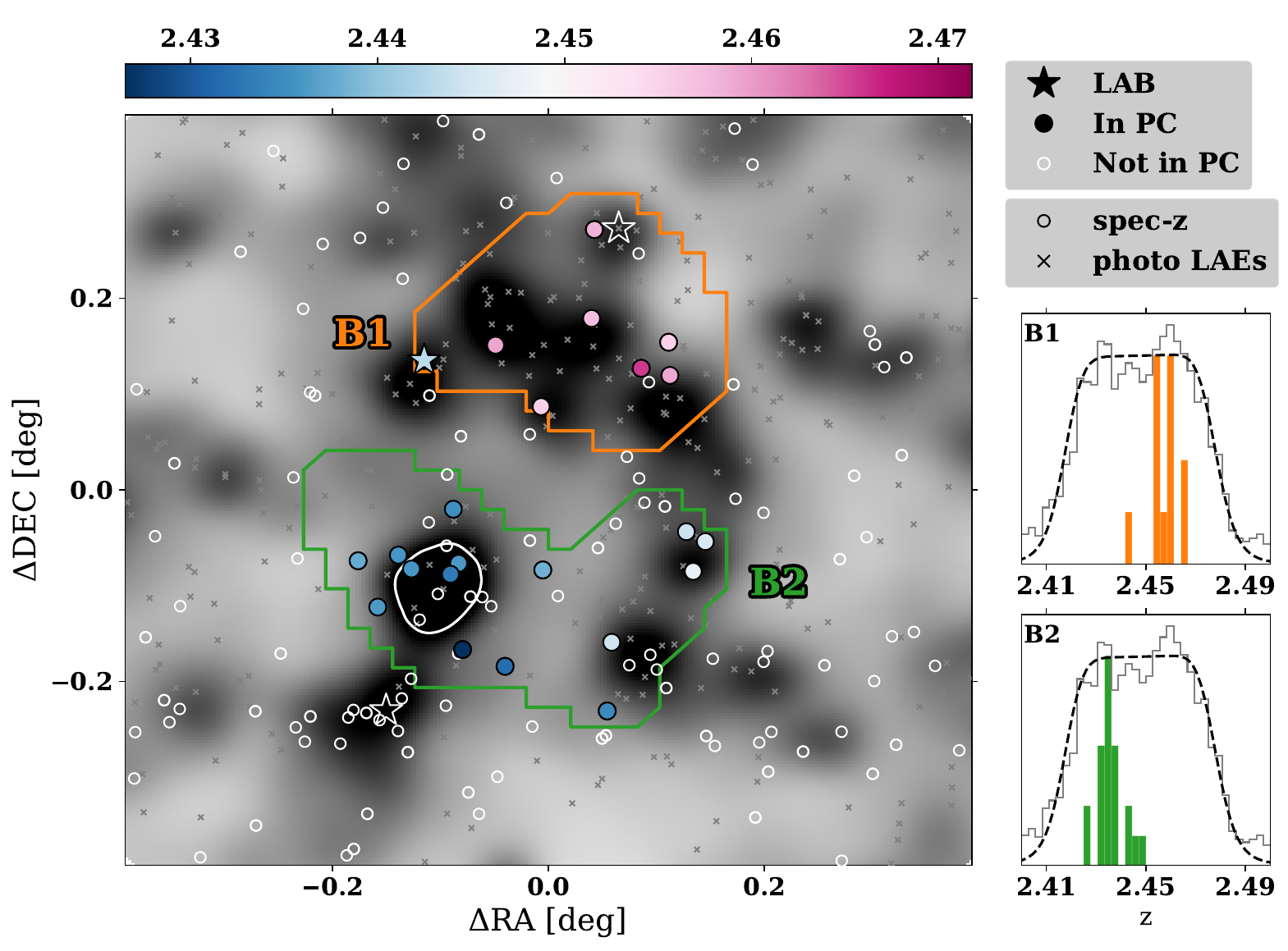}
    \end{minipage}%
    \hfill
    \begin{minipage}[c]{0.40\linewidth}
        \begin{interactive}{js}{COSMOS-z2.4-B-LAB.x3d}
            \includegraphics[width=\linewidth]{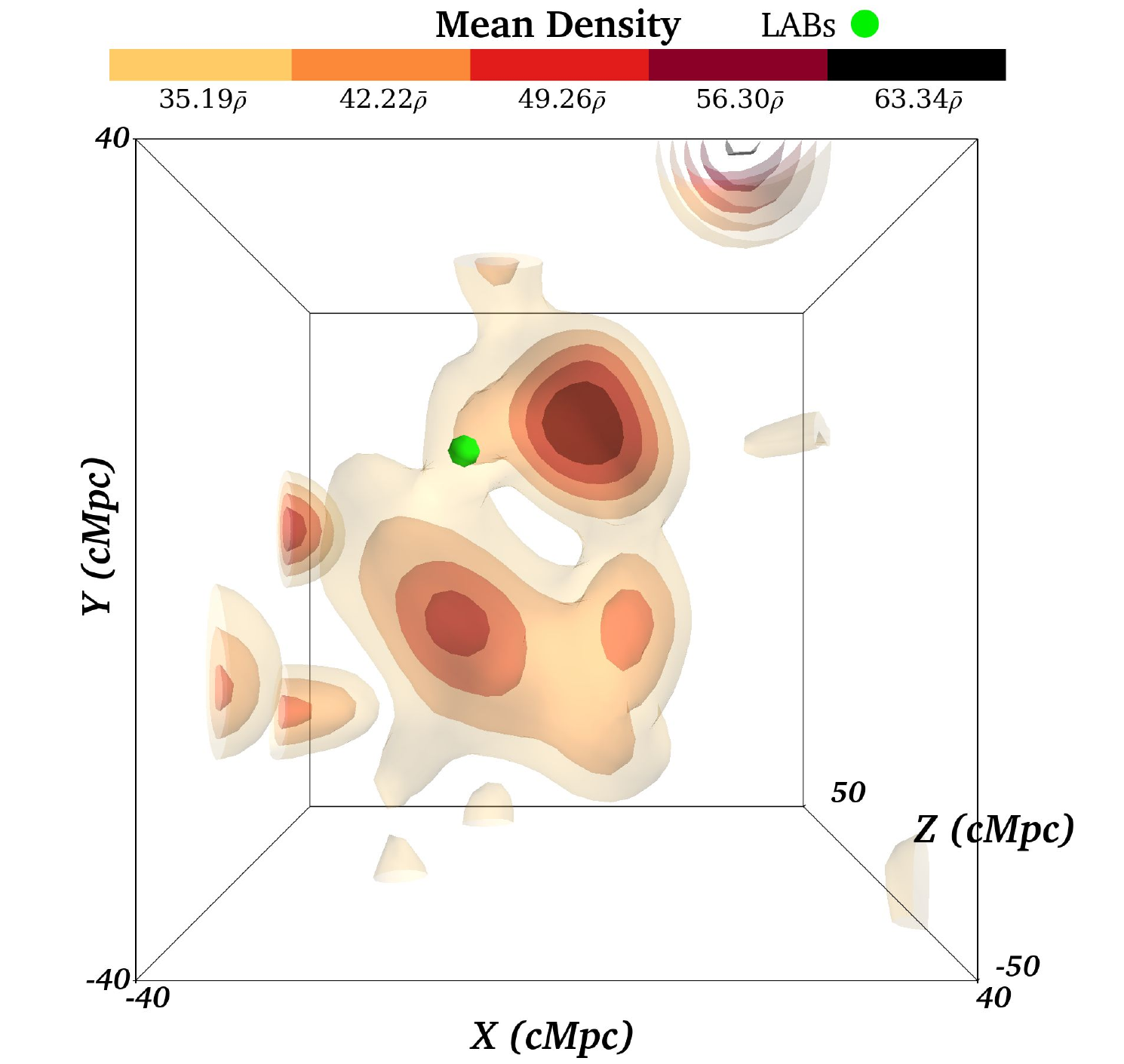}
        \end{interactive}
    \end{minipage}
    \caption{Similar to Figure~\ref{fig:cosmos_z3.1_B}, but for COSMOS-z2.4-B.}
    \label{fig:cosmos_z2.4_B}
\end{figure*}

Figure~\ref{fig:cosmos_z2.4_B} illustrates the structure as consisting of two groups with similar masses, estimated at $\log M_{\rm est}^{z=0}/M_\odot \approx 14.6\pm 0.2$ each (Table~\ref{tab:density_peaks}). Group~B1 is located to the north and is more distant from us than Group B2. In both the 2D and 3D density maps, B2 appears to contain two smaller subcomponents that, at the adopted 3D density threshold, merge into a single connected system. 
Our LAB selection identifies three blobs in this region \citep{moon2026b}, but only one is spectroscopically confirmed at $z\approx 2.445$. The connection between it and COSMOS-z2.4-B remains undetermined because of the absence of spectroscopy in the region. 


\subsection{XMM-z2.4-A}
\begin{figure*}
    \centering
    \textbf{XMM-z2.4-A}\par\medskip
    \begin{minipage}[c]{0.57\linewidth}
        \includegraphics[width=\linewidth]{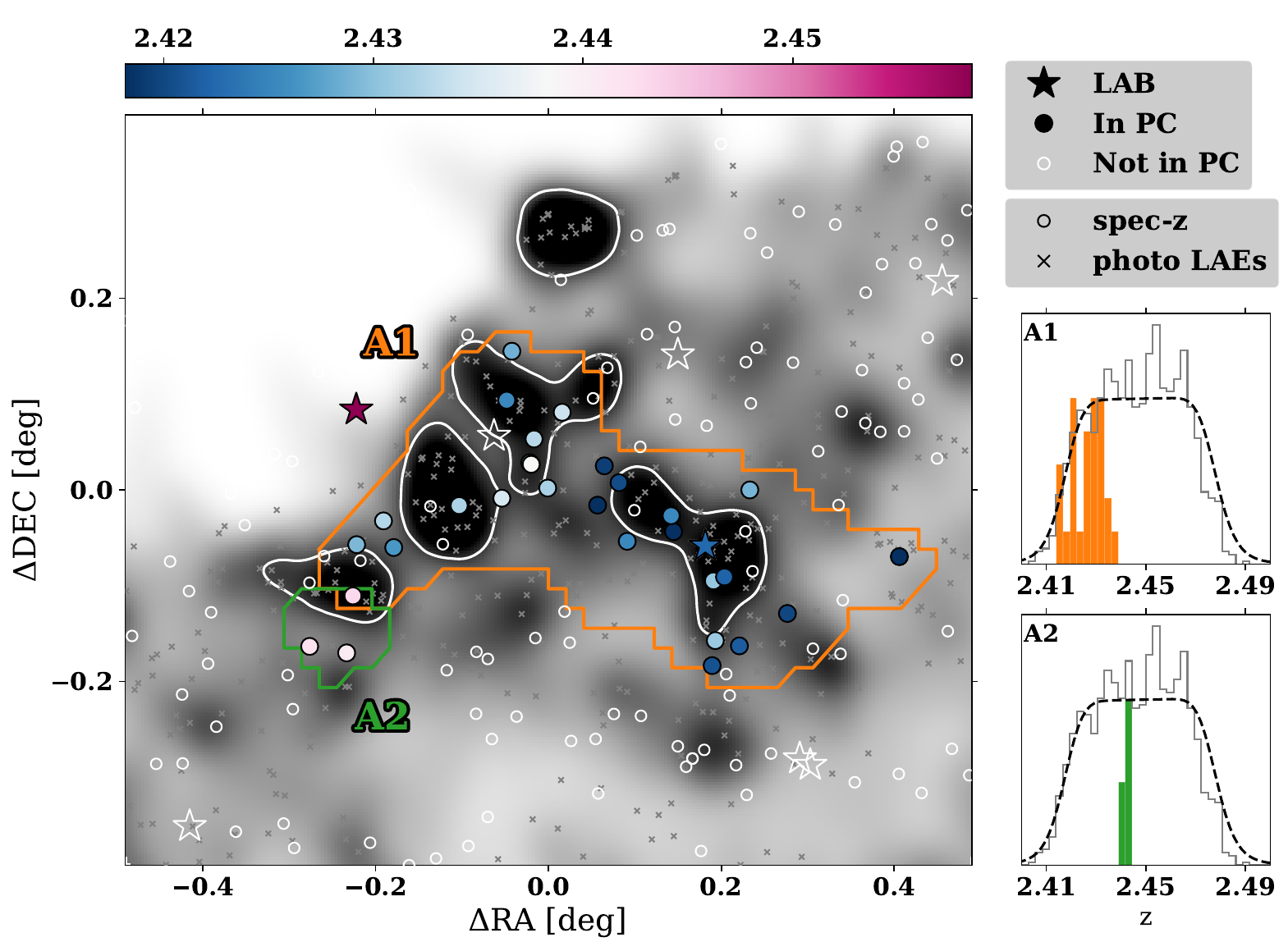}
    \end{minipage}%
    \hfill
    \begin{minipage}[c]{0.41\linewidth}
        \begin{interactive}{js}{XMM-z2.4-A-LAB.x3d}
            \includegraphics[width=\linewidth]{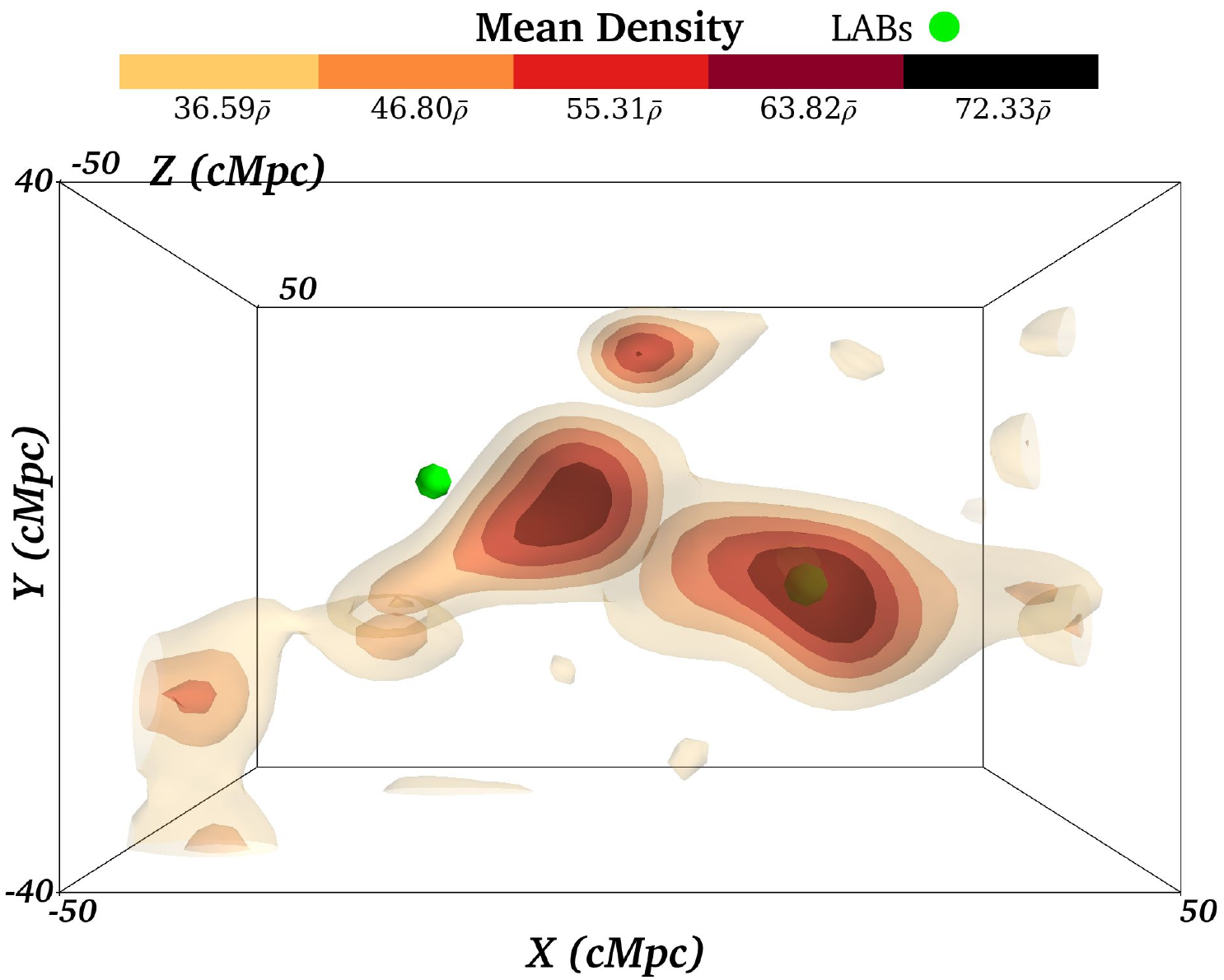}
        \end{interactive}
    \end{minipage}
    \caption{Similar to Figure~\ref{fig:cosmos_z3.1_B} but for XMM-z2.4-A.}
    \label{fig:xmm_z2.4_A}
\end{figure*}

Although XMM-z2.4-A lies near the northern boundary of the ODIN footprint, DESI spectroscopic coverage extends beyond this region. The full spatial extent of the structure beyond the surveyed area remains unconstrained. Figure~\ref{fig:xmm_z2.4_A} illustrates five distinct 2D overdensities (marked by white contours). The 3D reconstruction identifies four of these as interconnected systems at $z\approx 2.425$, while the fifth lacks spectroscopic data. Despite our detection picking up two distinct systems, group A2 is likely at the tip of A1 as it curves away from our viewpoint, a configuration that would be difficult to infer from projected density maps alone. With $\log M_{\rm est}^{z=0}/M_\odot \gtrsim 15.2$, XMM-z2.4-A stands as one of the most massive systems found in this study. Within this subfield, two LABs have been confirmed: one is situated farther back at $z\approx 2.46$, and the other resides near the core of A1. The latter is unusual because, as \citet{ramakrishnan23} reported, LABs tend to be found at the outskirts of protoclusters. 


\subsection{XMM-z2.4-B}

\begin{figure*}
    \centering
    \textbf{XMM-z2.4-B}\par\medskip
    \begin{minipage}[c]{0.57\linewidth}
        \includegraphics[width=\linewidth]{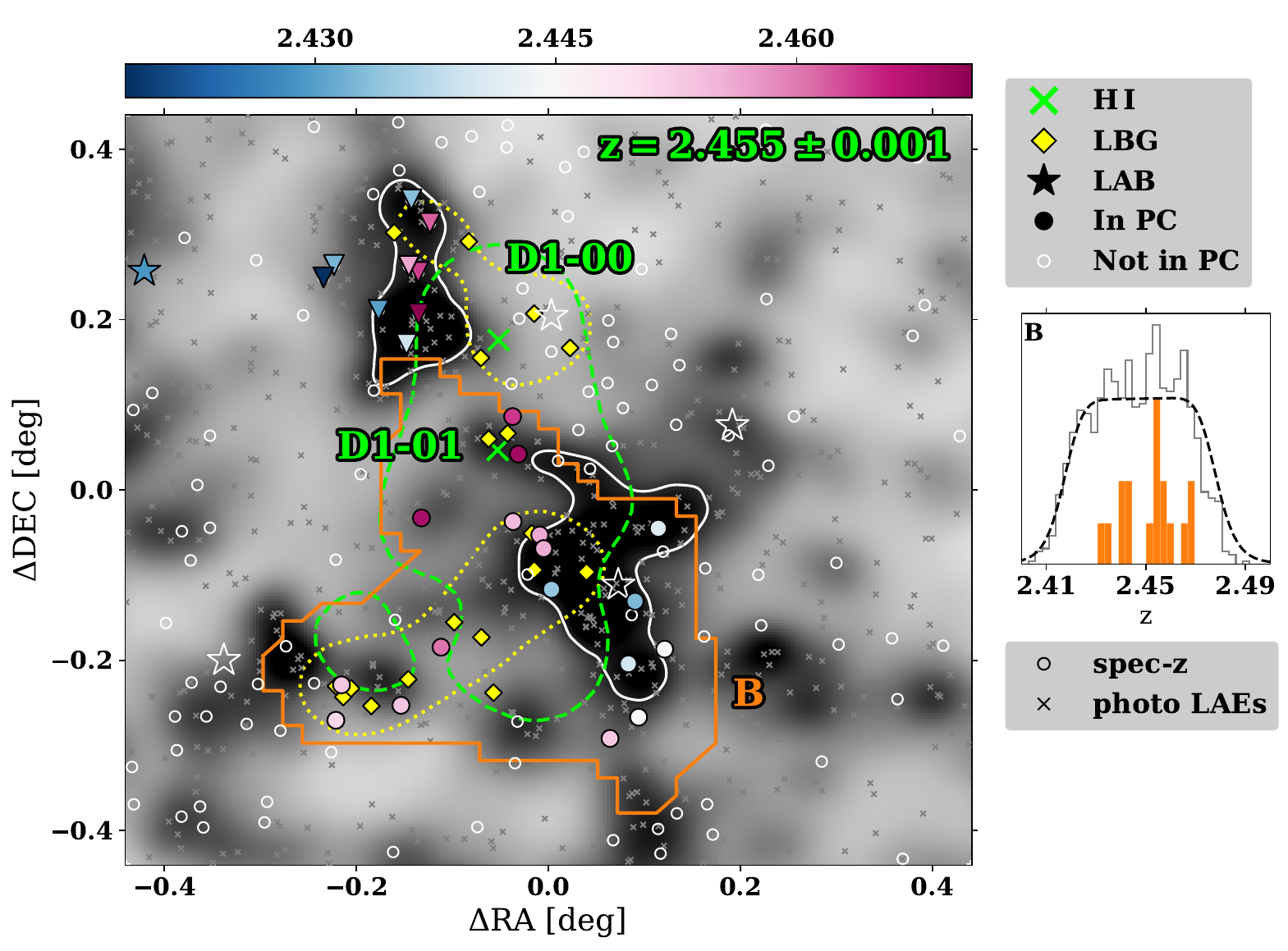}
    \end{minipage}%
    \hfill
    \begin{minipage}[c]{0.41\linewidth}
        \begin{interactive}{js}{XMM-z2.4-B-LAB.x3d}
            \includegraphics[width=\linewidth]{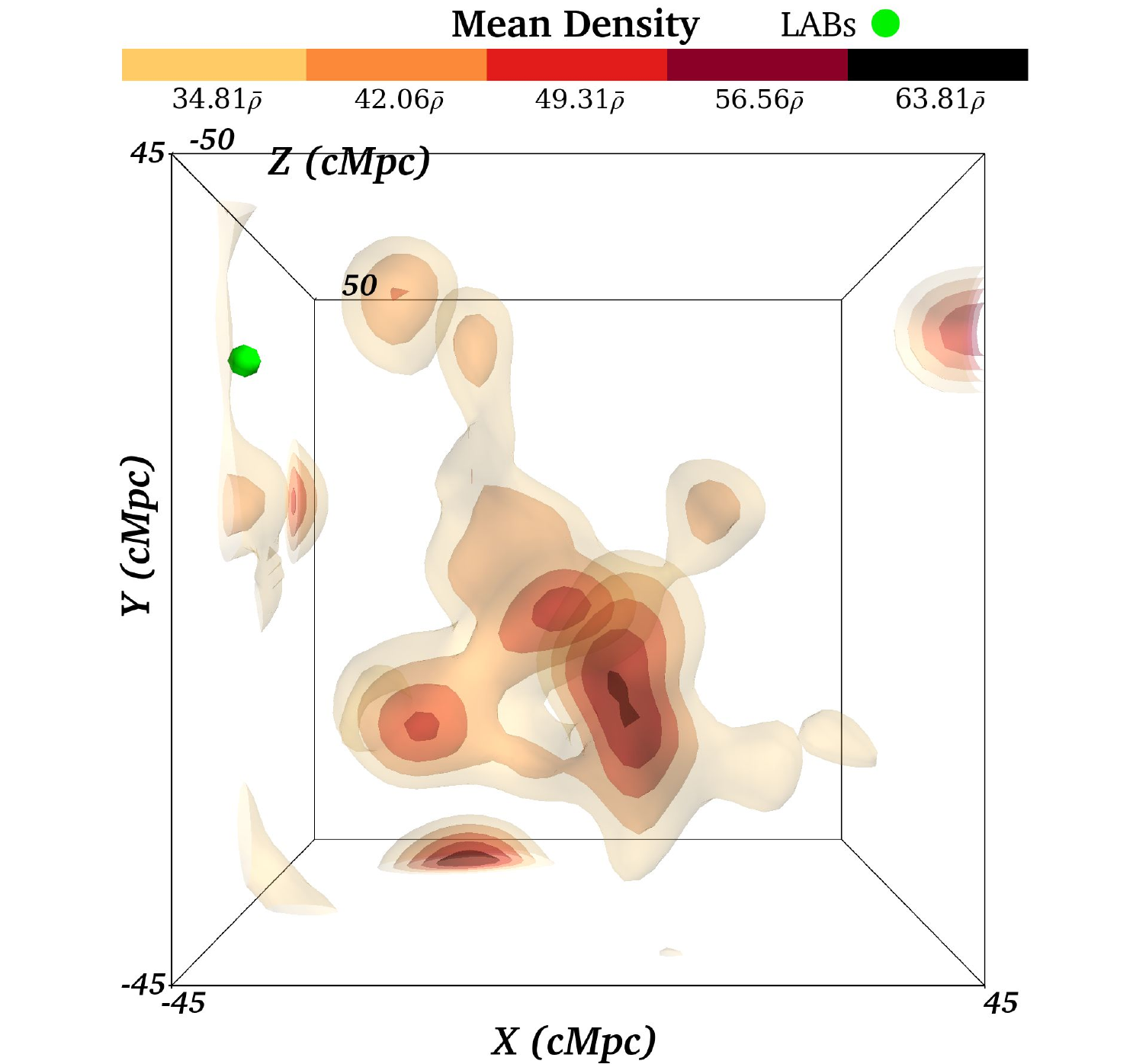}
        \end{interactive}
    \end{minipage}
    \begin{minipage}[c]{0.57\linewidth}
        \includegraphics[width=\linewidth]{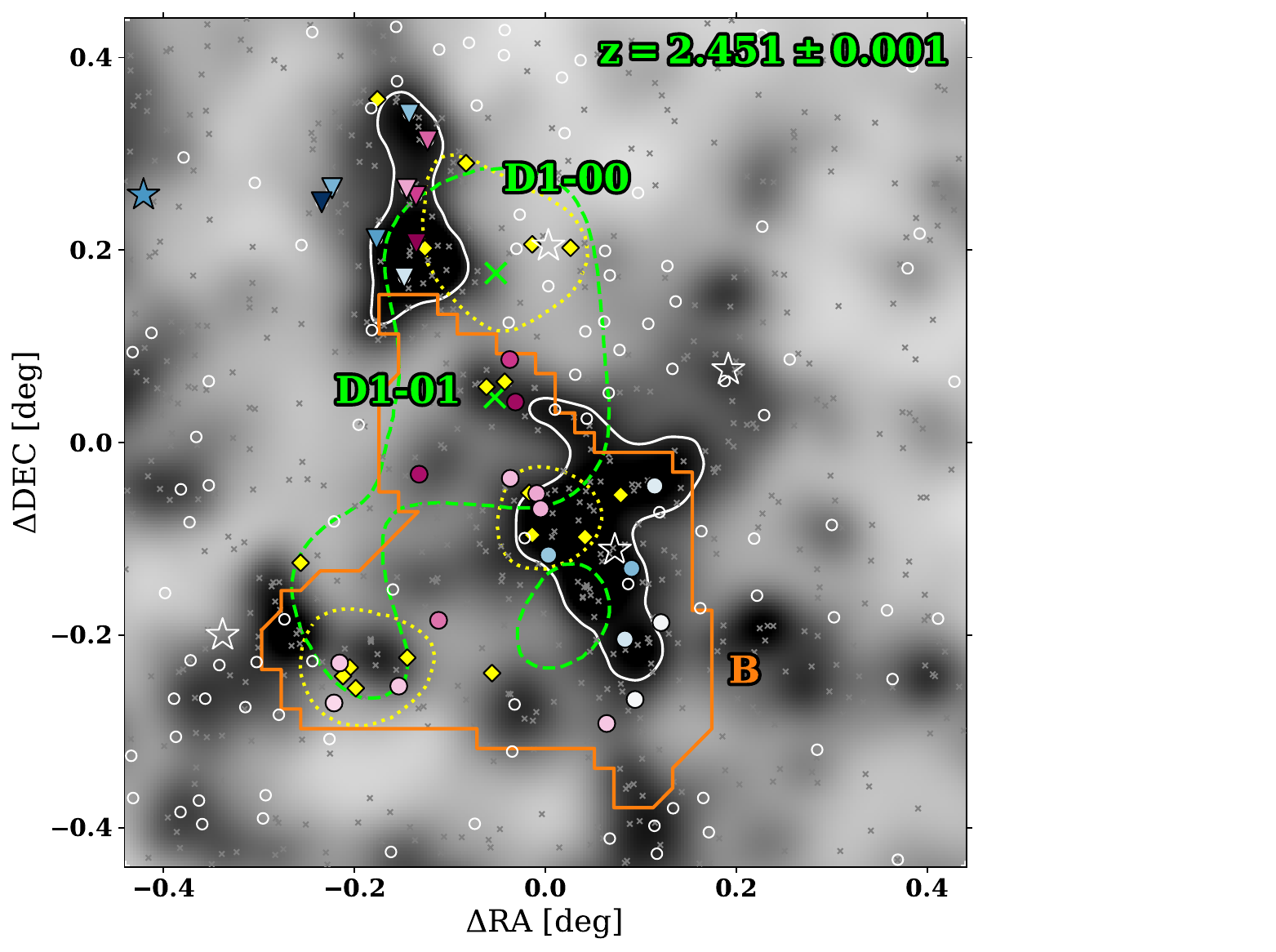}
    \end{minipage}%
    \hfill
    \begin{minipage}[c]{0.42\linewidth}
        \includegraphics[width=\linewidth]{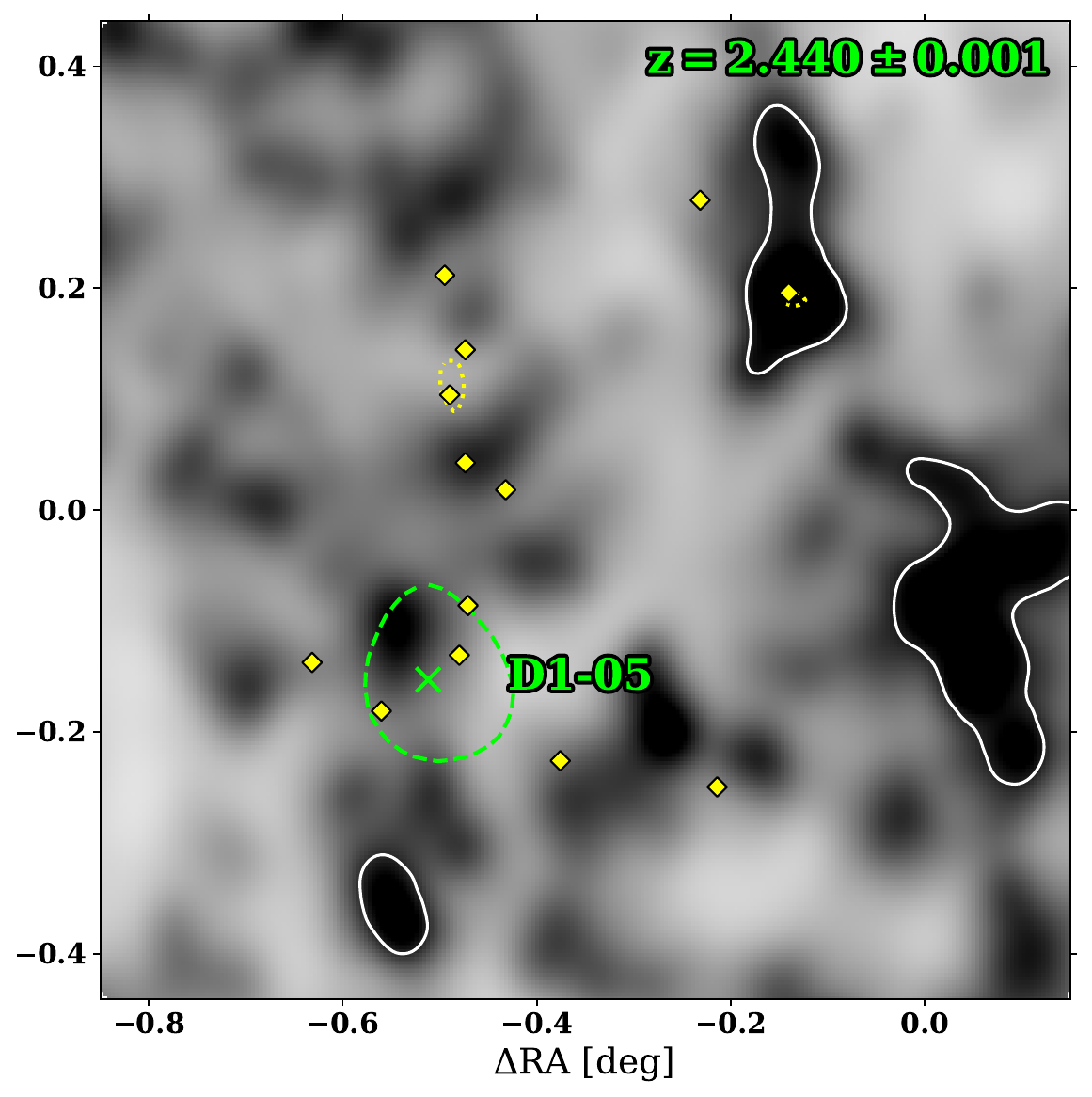}
    \end{minipage}
    \caption{Similar to Figure~\ref{fig:cosmos_z3.1_B} but for XMM-z2.4-B. In the top left, bottom left, and bottom right panel, green dashed contours mark the LATIS \lya absorption isosurface at $\delta_{\rm F}/\sigma_{\rm map}=-2$ (see Footnote~\ref{ft:delta_f} for the definition of this quantity), reported by \citet{newman25a}. These contours trace regions of enhanced H~{\sc i} absorption around the highest-significance LATIS peaks identified in their analysis, in three redshift slices corresponding to D1-00, D1-01, and D1-05, respectively, whose centers are indicated by green crosses. In the bottom right panel, the field center is shifted by $\approx$0.5~deg eastward to show D1-05. Yellow dotted contours mark the projected LATIS LBG overdensity contour at $\delta_{\rm LBG}/\sigma=2$, and yellow diamonds indicate individual LBG/QSO positions at these redshift slices from LATIS. The background greyscale is the 2D LAE surface density map. The downward triangles in the two left panels mark spectroscopic sources overlapping with D1-00; while these sources are not formally part of XMM-z2.4-B in our 3D detection, they reside in a 2D-selected protocluster whose angular extent overlaps significantly with D1-00 (see text for more discussion).
    }
    \label{fig:xmm_z2.4_B}
\end{figure*}

XMM-z2.4-B lies within the footprint of the Ly$\alpha$ Tomography IMACS Survey \citep[LATIS;][]{newman20}. LATIS uses background QSOs and galaxies to map the H~{\sc i} opacity of the intergalactic medium at $z=2.2–2.8$, based on the Ly$\alpha$ forest region of the spectra (i.e., $\lambda_{\rm rest}\sim1026$–1200~\AA), achieving a spatial resolution of 3–4~cMpc. Within a $\approx$0.6~deg$^2$ subregion of the XMM-LSS field, \citet{newman25a} identified seven protoclusters. Three of these fall within the redshift range probed by the $N419$ filter -- LATIS2\_D1\_00, LATIS2\_D1\_01, and LATIS2\_D1\_05—at $z=2.452$, 2.455, and 2.440, respectively (see their Table~1). Among the 37 density peaks identified across the full 1.7~deg$^2$ LATIS survey area, these rank 3rd, 7th, and 30th, respectively, in terms of the measured Ly$\alpha$ transmission deficit\footnote{Specifically, the ranking is based on $\delta_{\rm F}/\sigma_{\rm map}$, where $\delta_{\rm F}$ is the Ly$\alpha$ transmission fluctuation at a given 3D location, defined as $\delta_{\rm F}\equiv F/\langle F\rangle - 1$. Here $F$ is the measured transmitted Ly$\alpha$ flux, $\langle F \rangle$ is the mean transmission, and $\sigma_{\rm map}$ is the standard deviation of $\delta_{\rm F}$ across the field.\label{ft:delta_f}}, indicating that they correspond to regions of relatively high H~{\sc i} density in that order.

In Figure~\ref{fig:xmm_z2.4_B}, we present the projected 3D contours of XMM-z2.4-B (orange solid line) alongside the $\delta_{\rm F}/\sigma_{\rm map}$ contours from LATIS (green dashed lines), LBG surface-density contours (yellow dotted lines), and the positions of individual LBGs and QSOs from LATIS (diamonds). Two redshift slices are shown: $z=2.455\pm0.001$ (top left) and $z=2.451\pm0.001$ (bottom left). The centers of H~{\sc i} overdensities are indicated by large green crosses, labeled with their abbreviated LATIS IDs. Among these, LATIS2\_D1\_01 exhibits substantial—though not perfect—overlap with XMM-z2.4-B in both redshift and projected extent. Notably, its peak position (green cross) is slightly offset eastward from the region of highest LAE density, shown by the white contour. A similar spatial relationship between LAEs, LBGs, and H~{\sc i} overdensities has been noted in previous \lya forest tomography work \citep{momose21, newman25a}, where LAEs tend to be offset from the highest H~{\sc i} peaks traced by LBGs.

The sky position of LATIS2\_D1\_00 lies near another LAE overdensity to the north, which is not formally identified as a protocluster by our selection method and is not part of XMM-z2.4-B. Spectroscopically confirmed galaxies in this northern region are shown as downward-pointing triangles in the left panels of Figure~\ref{fig:xmm_z2.4_B}. None of the LAEs in the highest-density area near the center of LATIS2\_D1\_00 have spectroscopic confirmation, while those with measured redshifts are roughly evenly divided between $z\approx2.425$ and $z\approx2.455$–2.460. This redshift split, combined with the limited number of confirmed sources, likely led us to miss any structures that may exist in this area. Although additional spectroscopy around LATIS2\_D1\_00 is required, it is plausible that it is associated with a yet-to-be-confirmed LAE-rich structure.

In the bottom right panel of Figure~\ref{fig:xmm_z2.4_B}, we show the LAE density map around the third H~{\sc i}-selected overdensity, LATIS2\_D1\_05, located $\sim$0.5$^\circ$ east of XMM-z2.4-B. We do not detect any significant 3D structure in this region, although it may coincide with a relatively modest LAE overdensity. This is consistent with the relatively low ranking of LATIS2\_D1\_05 (30th out of 37 H~{\sc i}-selected overdensities), suggesting that it traces a much smaller and less prominent structure compared to the other two.


\subsection{XMM-z3.1-A}

\begin{figure*}
    \centering
    \textbf{XMM-z3.1-A}\par\medskip
    \begin{minipage}[c]{0.57\linewidth}
        \includegraphics[width=\linewidth]{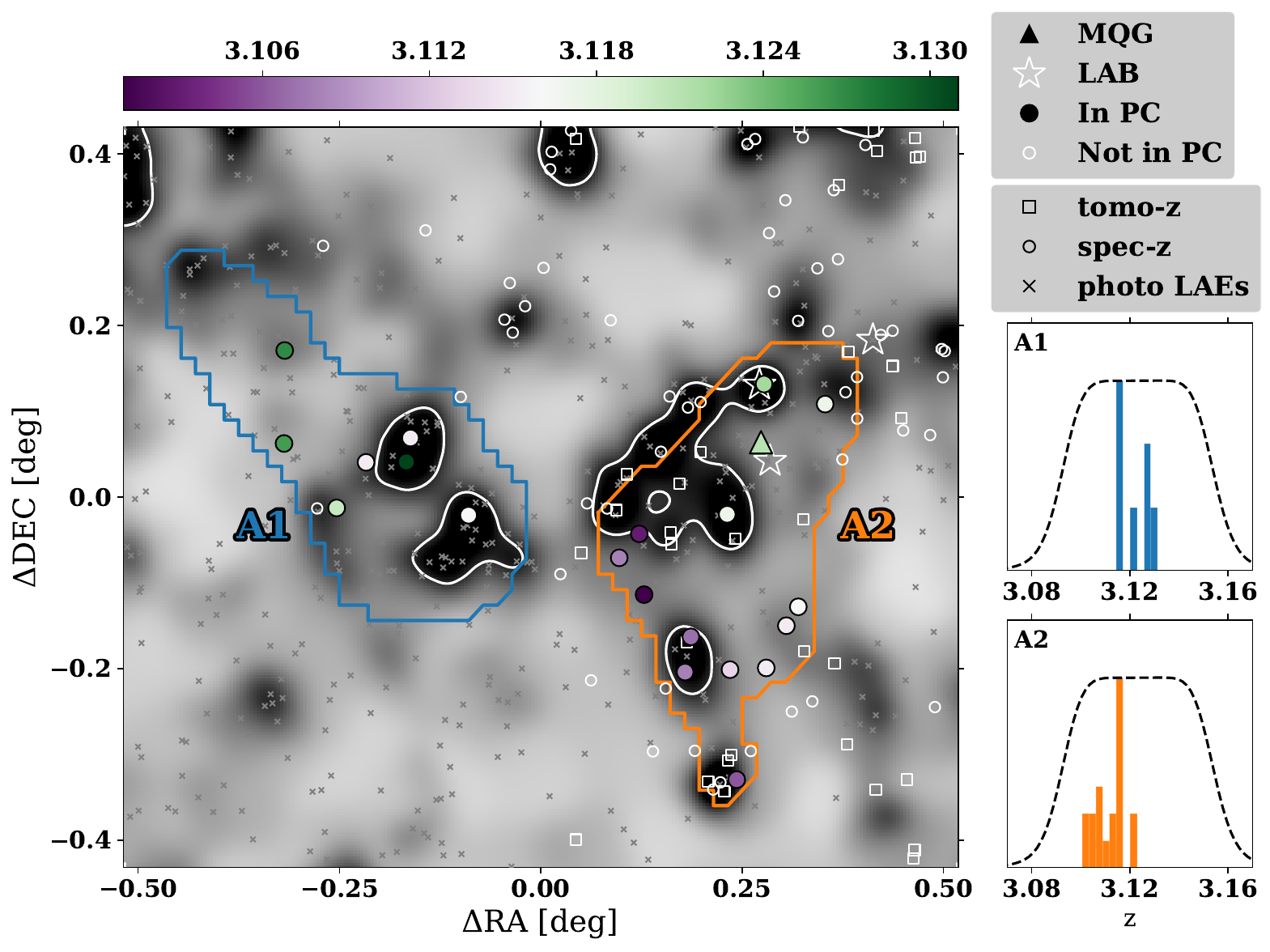}
    \end{minipage}%
    \hfill
    \begin{minipage}[c]{0.41\linewidth}
        \begin{interactive}{js}{XMM-z3.1-A-MQG.x3d}
            \includegraphics[width=\linewidth]{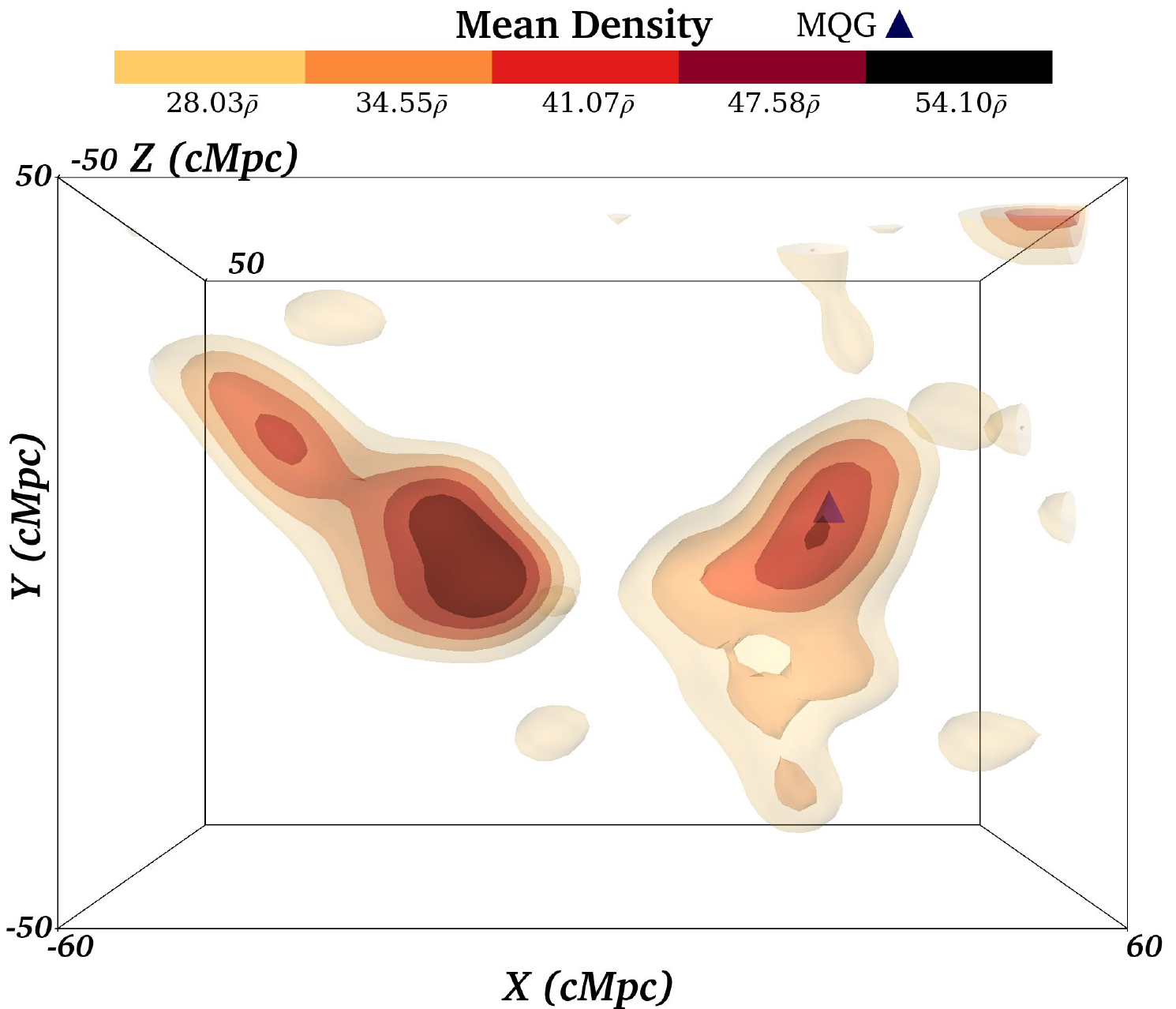}
        \end{interactive}
    \end{minipage}
    \caption{As Figure \ref{fig:cosmos_z3.1_B}, for XMM-z3.1-A. 3D-UDS-41232, a massive quiescent galaxy at $z=3.121$ \citep{nanayakkara25} is shown as a triangle. \emph{Left:} LAEs with redshifts spectroscopically confirmed are shown as circles, while redshifts estimated through tomography are shown in squares.}
    \label{fig:xmm_z3.1_A}
\end{figure*}

The location of XMM-z3.1-A lies in a part of the field that is not covered by DESI-ODIN spectroscopy since DESI has allocated it for different targets. We combine follow-up spectroscopy from \citetalias{Guo2020} alongside tomographic redshift estimates, as detailed in Section~\ref{sec:tomo}, to perform 3D detection and analysis of XMM-z3.1-A. 
Since the majority of objects have tomographic redshifts rather than spectroscopic ones, the effective line-of-sight resolution in our reconstruction is set by the tomography scatter (see Section~\ref{sec:tomo}).
In Figure~\ref{fig:xmm_z3.1_A}, these redshifts are represented by various symbols.

We identify two structures with descendant masses of similar magnitude ($\log M_{\rm est}^{z=0}/M_\odot\approx 15.1-15.2$). Together, they include four separate LAE surface overdensities shown as white contours. The boundaries of these structures are merely 2--3\arcmin\ apart. Structure A1 seems to have a dense core, with its tail extending towards the northeast, although further spectroscopy is required for confirmation. Meanwhile, the densest part of A2 is situated at its northern extremity, with the tail extending linearly southward, curving slightly in our direction. Three LABs have been identified, all near the northern area, though none have been spectroscopically confirmed.

Located at the northern end of A2 is a massive, quiescent galaxy (MQG) known as 3D-UDS-41232, which is among the 19 massive galaxies confirmed by \citet{nanayakkara25}. UDS-41232 ranks second in their list in stellar mass with $M_{\rm star}\approx 1.2\times 10^{11}M_\odot$, and its spectral energy distribution is consistent with a passive galaxy, with the time-averaged star formation rate over the past 100 million years being much less than $0.1M_\odot~{\rm yr}^{-1}$. Intriguingly, 3D-UDS-41232 is situated just outside the region of the highest LAE surface density. One LAB lies close to this MQG in projection, though the lack of its redshift prevents us from assessing any potential physical association between the two. 

According to \citet{nanayakkara25}, this galaxy assembled $\sim50\%$ of its stellar mass by $z\sim4.2$ and quenched by $z\sim3.5$, implying a rapid formation and quenching timescale of $\sim0.35$~Gyr. The relatively recent quenching is also consistent with the presence of residual H$\alpha$ emission in its spectrum. While further spectroscopy is necessary for a detailed study, this system represents one of the first ultramassive quiescent galaxies identified within a well-characterized LSS. Although the presence of a single MQG does not by itself demonstrate enhanced quenching in dense environments, its rapid assembly and recent quenching are consistent with scenarios in which galaxy evolution may be accelerated in overdense regions.



\section{Ly$\alpha$ Line Flux as a function of environment}\label{sec:lya_properties}

In dense protocluster environments, Ly$\alpha$ line fluxes may vary for different reasons. First, a steadier and robust supply of pristine gas may lead to a higher star formation rate, star formation efficiency, and/or younger stellar ages, thereby increasing Ly$\alpha$ photon production relative to the average environment. However, a higher star formation activity is expected to be accompanied by increased production of dust, which might offset or even reverse the trend. Second, regions of large-scale galaxy overdensities are expected to be also overdense with H~{\sc i} gas as they correlate with each other \citep[e.g.,][]{liang21,momose21,newman25b} although opposite examples also exist \citep{liang25}. In this context, the radiative transport process in protoclusters may be such that Ly$\alpha$ photons in their cores have shorter mean free paths and a higher likelihood of being 
absorbed or scattered \citep[e.g.,][]{Shimakawa2017,momose21}. 

Indeed, previous studies reported different Ly$\alpha$ properties in protocluster galaxies, but their conclusions did not always agree \citep[e.g.,][]{Shi2019,lemaux22}. Based on two confirmed protoclusters, \citet{Dey2016} reported that Ly$\alpha$ fluxes in protocluster galaxies are on average $\approx$35\% higher than non-protocluster counterparts, hinting that the effect may be non-negligible. Since both target selection and spectroscopic confirmation are based on the line brightness, this `luminosity bias' will impact both at the protocluster scales. More recently, \citet{nagaraj25} measured Ly$\alpha$ luminosity functions at $z\sim 2.4$, 3.1, and 4.5 based on ODIN LAE samples, finding that galaxies identified in 2D-selected protocluster candidates \citep{Ramakrishnan24} have shallower faint-end slopes $\alpha$ relative to the average field galaxies. 

Related trends have also been identified in simulations. Using TNG100 and TNG300, \citet{andrews25} found an excess of bright UV-selected galaxies in protocluster regions and a noticeably flatter faint-end slope in the UV luminosity function compared to the field, while the corresponding effect in the \lya luminosity function was present but significantly weaker.

Leveraging the robust sample of protocluster member galaxies identified in this study, we analyze the differences in line flux between LAEs found in average environments and those in protocluster environments. 
 We measure Ly$\alpha$ line fluxes by fitting each spectrum within $\pm40$~\AA\ of the central wavelength of the narrowband filter, assuming a two-component model with a linear continuum and a skew-normal function\footnote{A skew-normal function is defined as a normal distribution with an additional skew parameter $\alpha$ which controls the symmetry of the curve.} to capture the Ly$\alpha$ emission line. Once the best-fit is obtained, the line flux is estimated by simply integrating the skew-normal function. Given the limited depth of the spectroscopic observations, for the majority of the LAEs, the continuum fluxes are consistent with zero within the uncertainties. The best-fit continuum model does not change the line flux. 

    \label{fig:line_flux_ew}

\begin{figure*}
    \centering
    \includegraphics[width=0.9\linewidth]{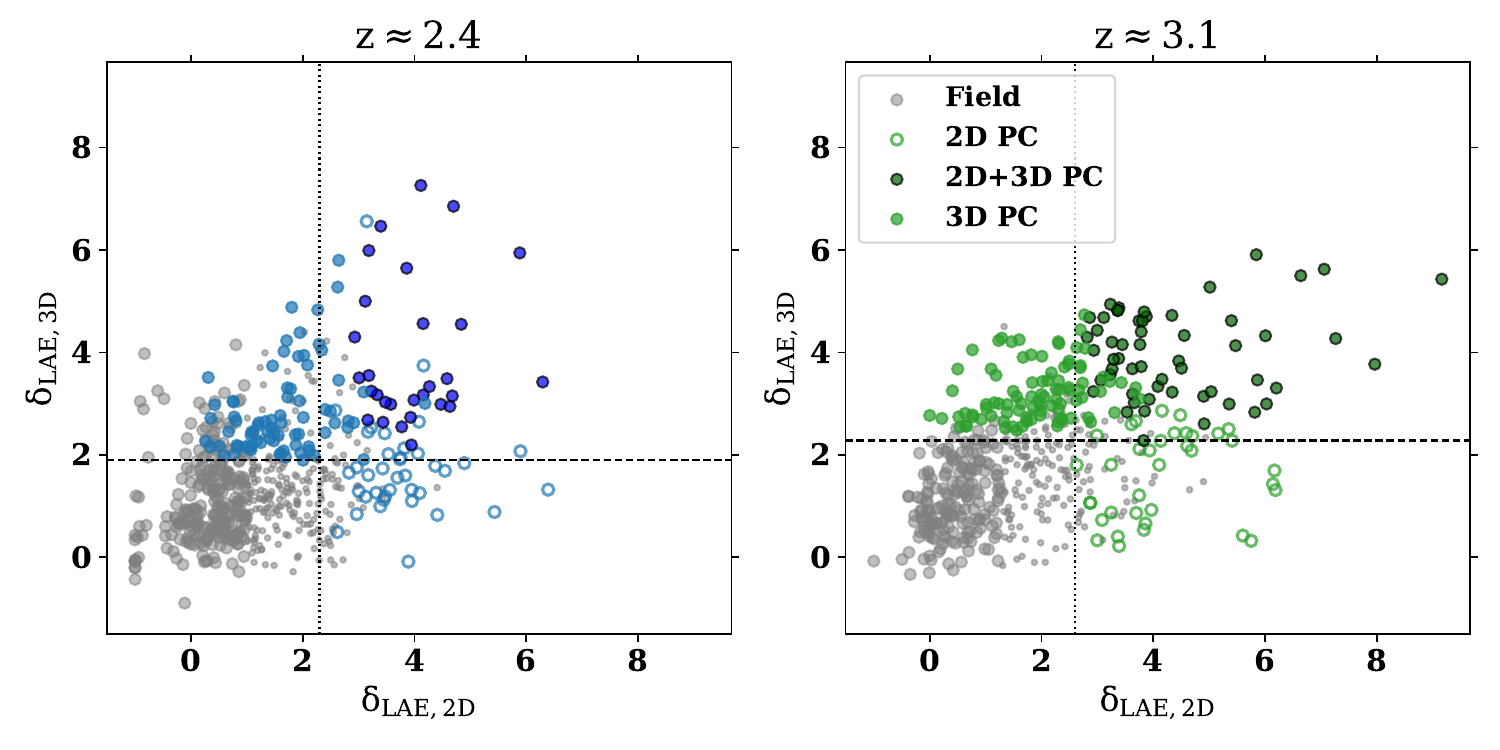}
    \caption{Comparison of overdensities in 2D and 3D space for LAEs at $z\approx2.4$ (left) and $z\approx3.1$ (right). Each point indicates an individual LAE, color-coded by its protocluster classification. Open (filled) circles denote members of the 2D (3D) sample only, and filled circles with a black outline denote members of both 2D and 3D protoclusters. Grey circles are field galaxies, while smaller grey dots are galaxies not in any sample. The dashed line marks the corresponding $\delta_{\rm LAE,3D}$ threshold used to define 3D protoclusters, and the dotted line marks the $\delta_{\rm LAE,2D}$ threshold.}
    \label{fig:delta_LAE}
\end{figure*}

\begin{figure*}
    \centering
    \includegraphics[width=\linewidth]{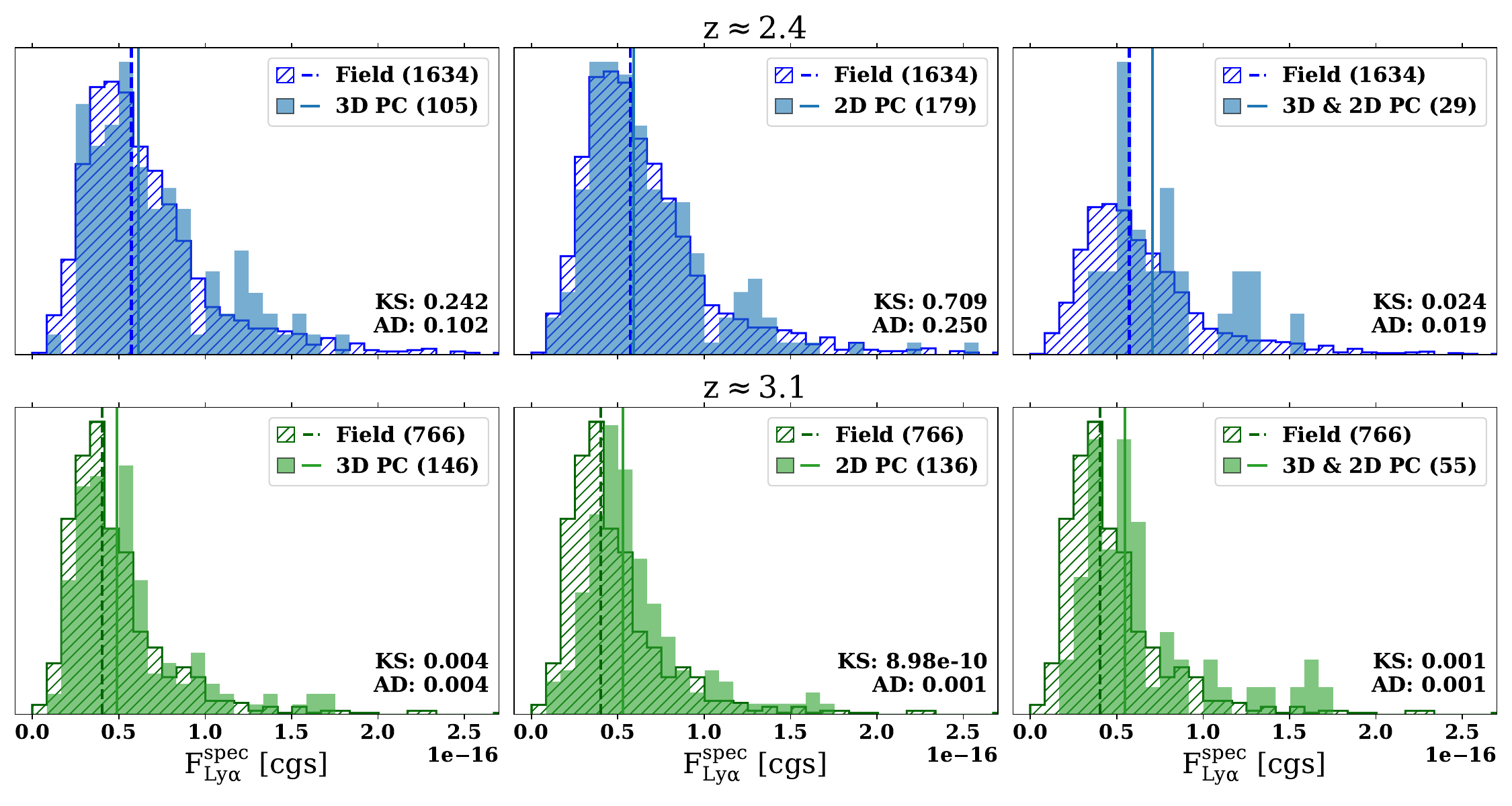}
    \caption{Histograms of $F_{{\rm Ly}\alpha}^{\rm spec}$ for protocluster (filled) and field (hatched) samples at $z \approx 2.4$ (top) and  $z \approx 3.1$ (bottom). The left (middle) panels show 3D (2D) protocluster member galaxies, while the right panels show galaxies identified as members by both 2D and 3D selection. Field galaxies are chosen to have local density $\delta_{\rm LAE,2D}$ at the 50\%-tile or below. Vertical solid (dashed) lines mark the median line fluxes of the protocluster (field) sample. The Kolmogorov-Smirnov (KS) and Anderson-Darling (AD) test $p$-values, comparing the two distributions, are listed in the bottom right of each panel.}
    \label{fig:lya_histogram_flux}
\end{figure*}


We examine how the large-scale environment affects the Ly$\alpha$ line fluxes of galaxies. In this work, we use two operational definitions of protoclusters: the 2D protocluster selection is based on projected LAE surface density (see Section~\ref{sec:odin_data}), while the 3D protocluster selection is based on the reconstructed 3D LAE density field (Section~\ref{sec:desc_mass}). We therefore assess environmental trends using both the 2D and 3D protocluster samples.
For clarity, we reiterate the criteria here. 2D protoclusters are identified as contiguous regions with projected overdensity $\delta_{\rm LAE,2D} > 2.3$ (2.6) at $z\approx 2.4$ (3.1) and projected area $> 40$~cMpc$^2$. 3D protoclusters are defined as contiguous volumes in the reconstructed density field that exceed the 97th percentile overdensity and enclose $>500$~cMpc$^3$.

Figure~\ref{fig:delta_LAE} shows the spectroscopically confirmed LAEs used in this analysis, plotted in the $\delta_{\rm LAE,2D}$–$\delta_{\rm LAE,3D}$ parameter space. Galaxies identified as members of 2D (3D) protoclusters are marked by open (filled) circles. Field galaxies, shown as grey circles, are defined to have $\delta_{\rm LAE,2D}\leq 1.0$ (1.3) at $z\approx 2.4$ (3.1), which encloses the bottom half in 2D density. By doing so, we ensure that no galaxy near a 2D or 3D protocluster is included in the field sample. The rest, belonging to neither our 2D/3D protocluster sample nor the field sample, indicated by smaller grey dots in Figure~\ref{fig:delta_LAE}, are not used in the subsequent analysis.

The relatively large scatter between $\delta_{\rm LAE,2D}$ and $\delta_{\rm LAE,3D}$ is readily understood. Our 3D detection algorithm is tuned to identify the entire volume of an overdensity that is likely to collapse into a single halo by $z=0$, ensuring that the corresponding descendant mass estimate approximates $M_{200}^{z=0}$. As demonstrated in \citet{Ramakrishnan24}, this approach recovers the present-day halo mass to within 0.15~dex. Consequently, a galaxy can occupy a region of high 3D density while still lying outside the local 2D density peak projected on the sky.

This behavior is illustrated in Figures~3–8, where the reconstructed 3D protocluster boundaries are shown as colored contours and the 2D photometric protoclusters are outlined in white. In contrast, the galaxies within 2D-selected protoclusters should trace the highest surface densities but contain some interlopers -- located in the foreground or background relative to the true 3D structure -- which dilute intrinsic correlations. In this sense, galaxies classified as members of {\it both} 2D and 3D protoclusters most likely reside in the densest cores of forming clusters.

In Figure~\ref{fig:lya_histogram_flux}, we compare the Ly$\alpha$ line-flux distributions of galaxies residing in protoclusters identified by the 3D (left), 2D (middle), and combined 2D\&3D (right) methods with those of field galaxies. Each histogram is normalized to unit area for direct comparison. To test whether the protocluster and field samples are drawn from the same parent population, we apply both the Kolmogorov–Smirnov and Anderson–Darling tests; the resulting $p$-values are shown in the bottom right corner of each panel.

At $z\approx3.1$, protocluster LAEs exhibit systematically higher Ly$\alpha$ fluxes than field galaxies, with the difference being statistically significant. Visual inspection and formal tests both indicate that this enhancement is most pronounced in the 2D and 2D\&3D selections, while it appears weaker for the 3D case alone. This likely reflects the inclusion of galaxies near the edges of 3D protoclusters, where the local density is lower (see Figure~\ref{fig:delta_LAE}). This suggests that not every protocluster member has a higher \lya flux than most field galaxies.
Although both the 2D and 2D\&3D protocluster samples differ significantly from the field, the shift in median flux (vertical lines in Figure~\ref{fig:lya_histogram_flux}) is larger for the 2D\&3D subset, consistent with the expectation that projection effects in purely 2D selections dilute the trend.

At $z\approx2.4$, the difference between protocluster and field LAEs is less pronounced. The 3D-selected protoclusters show only a marginal trend, and the 2D sample shows little to no distinction from the field. This likely reflects increased dilution from galaxies that are not located in the densest protocluster regions: projection effects in the 2D selection and the inclusion of galaxies in lower-density protocluster outskirts in the 3D selection both act to reduce the contrast relative to the field. However, galaxies identified as 2D\&3D members -- those located in the densest protocluster cores -- still show a statistically significant difference ($p\approx0.02$), with an enhancement in median line flux comparable to that observed at $z\approx3.1$.

Although the sample size is limited, our results suggest that galaxies in dense protocluster cores have Ly$\alpha$ flux distributions that are systematically skewed toward brighter values and show a relative deficit of faint emitters. This finding aligns with the shallower faint-end slope ($\alpha$) of the Ly$\alpha$ luminosity function reported for (2D) protocluster regions by \citet{nagaraj25}, although they acknowledged that the trend is marginal and larger sample sizes are needed to validate the trend. Additionally, our results are fully consistent with those independently reported by \citet{uzsoy25b}, who also used the 2D-based protocluster selection \citep[also see][]{huang21}. Additional spectroscopy will be needed to confirm this trend.

\section{Summary}\label{sec:summary}

Building on \citet{ramakrishnan25b}, we extend the 3D reconstruction methodology in this work to identify and characterize massive protoclusters in the COSMOS and XMM-LSS fields. Our analysis combines a large sample of spectroscopically confirmed LAEs from DESI with an even larger catalog of photometric LAEs (Table~\ref{tab:datasets}). Our main findings are summarized as follows.

\begin{enumerate}
\item 
We confirm six massive protoclusters, including three newly identified systems and three previously reported structures. Five of these are confirmed using DESI–ODIN spectroscopy. The sixth structure is verified using a combination of spectroscopy from \citetalias{Guo2020} and tomographic redshifts derived from two overlapping filters, $NB497$ and $N501$ (Section~\ref{sec:tomo} and Figure~\ref{fig:nb_tomo}). 
The reconstructed maps also reveal multiple substructures interconnected by lower-density filaments, suggestive of early-stage assembly within the cosmic web.

\item 
Three of the identified systems overlap with structures detected through other tracers, offering insight into how different galaxy populations and intergalactic gas trace the underlying LSS (Section~\ref{sec:odin_structures}).  

– {\it COSMOS-z3.1-B:} Located within a region rich in LBGs \citep[][Figure~\ref{fig:cosmos_z3.1_B}]{toshikawa24}, the LAE and LBG surface density maps show a modest offset, which may be attributed to different selection effects of LAE- and LBG-based protocluster detections, or alternatively, the projection effect of physically unrelated cosmic structures. The only LBG structure with spectroscopic confirmation lies $\approx$70~cMpc in the foreground, and thus is unlikely to be associated with COSMOS-z3.1-B.
Follow-up spectroscopy is needed to clarify the physical connection of the remaining structures.  

– {\it XMM-z2.4-B:} This structure overlaps with H~{\sc i} tomographic mapping and LBG overdensities from LATIS, showing significant spatial alignment among LAE-, LBG-, and H~{\sc i}-selected high-density regions (Figure~\ref{fig:xmm_z2.4_B}).  

– {\it XMM-z3.1-A:} It is possibly the most massive structure identified in this study, as its estimated descendant mass rivals that of the Coma cluster. A spectroscopically confirmed massive ($M_{\rm star}\approx 10^{11.2}M_\odot$) quiescent galaxy is located in the central region of a structure (A2: Figure~\ref{fig:xmm_z3.1_A}). The presence of an MQG within the central region of this structure is consistent with the possibility that dense environments may contribute to accelerated galaxy assembly and quenching at $z\sim3$, in line with emerging observation and theoretical work.

\item 
Using both 2D- and 3D-defined protocluster membership (Section~\ref{sec:desc_mass}), we compare Ly$\alpha$ fluxes in protocluster and field galaxies. By combining 2D and 3D density information, we isolate galaxies residing in the densest protocluster cores. At $z\approx 3.1$, these galaxies exhibit higher median Ly$\alpha$ line fluxes and a deficit of faint emitters compared to the field. Consistent with \citet{uzsoy25b}, no significant difference is found for $z\approx 2.4$ galaxies when selected purely by 2D overdensity. However, the 2D\&3D-selected subset shows a comparable flux enhancement to that observed at $z\approx 3.1$, albeit with a smaller sample size. Overall, the contrast between protocluster and field populations is much stronger at $z\approx3.1$ than at $z\approx2.4$, suggesting possible redshift evolution in the environmental dependence of Ly$\alpha$ emission.

\item Of the six structures considered in this work, four are the progenitors of the most massive structures whose estimated descendant masses rival that of Coma at $\log(M^{z=0}/M_\odot) \gtrsim 15$ (Table \ref{tab:density_peaks}). Together with the two protoclusters reported by \citet{ramakrishnan25b} and {\it Hyperion}, ODIN has identified a total of seven Coma progenitors within a volume of $\sim 1.7 \times 10^{7}$~cMpc$^{3}$. The inferred number density is consistent with the theoretical expectation, demonstrating that ODIN is highly successful at identifying the high-redshift progenitors of the most massive structures. 

\end{enumerate}

Combined with \citet{ramakrishnan25b}, this work brings the total number of spectroscopically confirmed ODIN protoclusters to eight, providing one of the first wide-field, 3D-mapped samples of massive protoclusters across Cosmic Noon.  \\

\section*{Data Availability}
All data shown in figures are available on Zenodo (doi: \href{https://zenodo.org/records/18436376}{https://doi.org/10.5281/zenodo.18436376}). Interactive 3D visualizations are available on GitHub (\href{https://ortiz140.github.io/odin/}{https://ortiz140.github.io/odin/}).

\section*{Acknowledgments}
We thank Jun Toshikawa for helpful discussions. We are grateful to the anonymous referee for a careful reading of the manuscript and a helpful report. The authors acknowledge financial support from the U.S. National Science Foundation (NSF) under grant Nos. AST-2206705, AST-2408359, and from the Ross-Lynn Purdue Research Foundations. 
This work is based on observations at Cerro Tololo Inter-American Observatory, NSF’s NOIRLab (Prop. ID 2020B-0201; PI: K.-S. Lee), which is managed by the Association of Universities for Research in Astronomy under a cooperative agreement with the National Science Foundation.
M.C.A. and A.K. acknowledge financial support from ANID BASAL project FB210003 and ALMA fund with code 31220021.
C.G. acknowledges support from the National Science Foundation under grant AST-2408358.
L.G. gratefully acknowledges support from the FONDECYT regular project number 1230591, the ANID BASAL project FB210003, ANID - MILENIO - NCN2024\_112.
H.S.H. acknowledges the support of the National Research Foundation of Korea (NRF) grant funded by the Korea government (MSIT), NRF-2021R1A2C1094577, and Hyunsong Educational \& Cultural Foundation.
S.L. acknowledges support from the National Research Foundation of Korea (NRF) grant RS-2025-00573214, funded by the Korean government (MSIT).
Y.Y. and B.M. are supported by the Basic Science Research Program through the National Research Foundation of Korea funded by the Ministry of Science, ICT \& Future Planning (2019R1A2C4069803).
H. Song was supported by the National Research Foundation of Korea(NRF) grant funded by the Korea government (MSIT) (No. RS-2025-25442707).

This material is based upon work supported by the U.S. Department of Energy (DOE), Office of Science, Office of High-Energy Physics, under Contract No. DE–AC02–05CH11231, and by the National Energy Research Scientific Computing Center, a DOE Office of Science User Facility under the same contract. Additional support for DESI was provided by the NSF, Division of Astronomical Sciences under Contract No. AST-0950945 to the NSF’s National Optical-Infrared Astronomy Research Laboratory; the Science and Technology Facilities Council of the United Kingdom; the Gordon and Betty Moore Foundation; the Heising-Simons Foundation; the French Alternative Energies and Atomic Energy Commission (CEA); the National Council of Humanities, Science and Technology of Mexico (CONAHCYT); the Ministry of Science, Innovation and Universities of Spain (MICIU/AEI/10.13039/501100011033), and by the DESI Member Institutions: \url{https://www.desi.lbl.gov/collaborating-institutions}. Any opinions, findings, and conclusions or recommendations expressed in this material are those of the author(s) and do not necessarily reflect the views of the U. S. National Science Foundation, the U. S. Department of Energy, or any of the listed funding agencies.
The authors are honored to be permitted to conduct scientific research on I’oligam Du’ag (Kitt Peak), a mountain with particular significance to the Tohono O’odham Nation.

\bibliography{myrefs}{}
\bibliographystyle{aasjournal}

\end{document}